\documentclass[twocolumn,noshowpacs,preprintnumbers,amsmath,amssymb]{revtex4}

\usepackage{amsfonts}
\usepackage{amssymb}
\newtheorem{theo}{Theorem}
\newtheorem{lemma}{Lemma}
\newtheorem{cor}{Corollary}
\newtheorem{defi}{Definition}
\newtheorem{prop}{Proposition}

\newcommand{\proofend}{\raisebox{1.3mm}{%
\fbox{\begin{minipage}[b][0cm][b]{0cm}\end{minipage}}}}
\newenvironment{proof}[1][\hspace{-1mm}]{{\noindent\it Proof #1:}
}{\mbox{}\hfill \proofend\\\mbox{}}

\def\2{{\textstyle\frac{1}{2}}}
\def\ba{\begin{eqnarray}}
\def\ea{\end{eqnarray}}
\def\be{\begin{equation}}
\def\ee{\end{equation}}

\newcommand{\CL}{{\cal{L}}}
\newcommand{\CT}{{\cal{T}}}
\newcommand{\CP}{{\cal{P}}}

\newcommand{\CS}{{\cal{S}}}
\newcommand{\CX}{{\cal{X}}}
\newcommand{\CD}{{\cal{D}}}
\newcommand{\CB}{{\cal{B}}}
\newcommand{\CG}{{\cal{G}}}
\newcommand{\CH}{{\cal{H}}}
\newcommand{\CC}{{\cal{C}}}
\let\a=\alpha \let\b=\beta \let\g=\gamma

  \let\l=\lambda \let\m=\mu
\let\n=\nu   \let\r=\rho \let\s=\sigma
 \let\o=\omega  
   
\let\O=\Omega \let\S=\Sigma 
  \let \G=\Gamma

\def\({\left(} \def\){\right)} \def\<{\langle} \def\>{\rangle}

 \let\LRA=\Leftrightarrow

 \let\La=\Leftarrow

\newcommand*{\dR}{{\mathbb R}}

\newcommand{\md}{\mathrm{d}}

\usepackage{graphicx}
\usepackage{dcolumn}
\usepackage{bm}

\begin{document}


\title{Gravity from Lie algebroid morphisms}

\author{Thomas Strobl}
\altaffiliation[]{ITP FSU Jena, Max-Wienpl.~1, 07743 Jena, Germany}
 \email{pth@tpi.uni-jena.de}


\begin{abstract}
Inspired by the Poisson Sigma Model and its relation to 2d gravity,
we consider models governing
morphisms from $T\S$ to any Lie algebroid $E$, where $\S$ is regarded
as  $d$-dimensional spacetime manifold. We address the question of minimal
conditions to be placed on a bilinear expression in the
1-form fields, $\CS^{ij}(X) A_i A_j$, so as to
permit an interpretation as a metric on $\Sigma$. This becomes a
simple compatibility condition of the $E$-tensor $\CS$ with the chosen
Lie algebroid structure on $E$. For the standard Lie algebroid $E=TM$
the additional structure is identified with a Riemannian
foliation of $M$, in the Poisson case  $E=T^*M$ with a
sub-Riemannian structure which is Poisson invariant with
respect to its annihilator bundle.
(For integrable image of $\CS$, this means that the induced
Riemannian leaves should be invariant with respect to all
Hamiltonian vector fields of functions which are locally constant 
 on this foliation). 
This provides
a huge class of new gravity models in $d$ dimensions, embedding known
2d and 3d models as particular examples.

\end{abstract}

\maketitle

\section{\label{sec:level1}Introduction}

\subsection{The problem---in the 2d Poisson setting}
In the present paper we want to address essentially the following
question: Given a manifold $M$ equipped with
some 2-tensor $\CT$, say contravariant, and given a two-dimensional manifold
$\Sigma$, which we call worldsheet or spacetime,
under what conditions can we define a ``reasonable''
theory of gravity on $\Sigma$ with just these data? Certainly one
needs to specify what one means by reasonable, and we will do this---in a
rather minimalistic but also precise way---below. Moreover, we will
address a likewise question also for higher dimensional spacetime manifolds
$\S$; but the formulation of the respective higher dimensional
problem, including a replacement of $M$ by a vector bundle with
structure---which then also explains the title of the present
work---needs further background material, which we will present only
later on.

Without any further restriction, the above question is not very fruitful:
Indeed, for a generic choice of $\CT$, we may split $\CT$ into its
symmetric
and antisymmetric part, $\CS$ and $\CP$, respectively. Then we may
use the antisymmetric part to
define the following canonical action functional
\be L[X,A] \:  \colon \! \!\!
= \int_\S A_i \wedge \md X^i + \2 \CP^{ij} A_i \wedge
A_j \, ,
\label{action} \ee
where $X^i(x)$ parametrizes the map $\CX$ from $\Sigma$ to $M$, which
is used as a target manifold, and $A = A_i \otimes \md X^i \equiv
 A_{\mu i}(x) \md x^\mu \otimes
\md X^i$ is a 1-form on $\S$ with values in
the pullback bundle $\CX^*T^*M$. The symmetric part, on the other
hand, may be used to define a metric on $\Sigma$ by means of
\be g  \,  \colon \! \!\!= \; \2 \CS^{ij} A_i A_j \, .  \label{metric} \ee
(Throughout this paper any suppressed tensor symbol will denote a
symmetrized tensor product, $A_i A_j \equiv A_i \otimes A_j +
A_j \otimes A_i$, while the antisymmetrized one, the wedge
product, will be displayed explicitly).
By construction, the functional (\ref{action}) is
invariant with respect to diffeomorphisms of $\Sigma$, which clearly
is a necessary condition for a theory describing gravity. For generic
enough choice of $\CT$ (and thus also of $\CP$) there are no further local
or gauge symmetries of $L$ {\em and\/} there will be solutions to the
field equations 
\ba \md X^i + \CP^{ij}(X) A_j = 0
\label{eqs1} \\  \md A_i + \2
\CP^{kl}{},_{i}(X) \, A_k \wedge
A_l = 0 \label{eqs2} \ea 
such that $g$ constructed according to (\ref{metric}) satisfies $\det
g \neq 0$ on all of $\Sigma$. This latter condition is also 
necessary for regarding $g$ as a metric tensor field. We may, however,
not expect it to hold for {\em all\/} solutions to the field
equations: For example, $A_i :\equiv 0$, $X^i :\equiv const$ is a
solution to the field equations (\ref{eqs1}), (\ref{eqs2}), leading to
$g \equiv 0$ irrespective of whatever $\CT$ is. Still, $\CS$
should not be e.g.\ vanishing, since then there would be {\em no\/} 
gravitationally interesting or admissible solutions; for a generic
enough choice of $\CT$ this will not happen, however, and then there
is no obstacle to regard (\ref{action}) and (\ref{metric}) as defining
some 2d theory of gravity. 

Note that we did not require $L$ to be a functional of $g$
alone or a functional of $g$ at all.  Applications, some of which
we will recall in the body of the paper, show that this
would be too restrictive: There may be some other ``parent'' action
functional, which depends explicitly on $g$, or maybe only on some
vielbeins from which $g$ results by the standard bilinear combination,
and possibly on some other
more or less physical fields and only after some (possibly involved)
identifications, including (\ref{metric}), one ends up with the
functional (\ref{action}). 

The initially posed question bears 
more structure\footnote{This is true at least from the mathematical
point of view. The condition imposed excludes propagating
modes, which is not so desirable from the physical point of
view. Still, many examples in two dimensions have this feature, and many
more theories may be captured if the topological theory describes only
one sector of a more extended one, cf.~e.g.~\cite{Habil2,Kummerreview}
as well as the discussion further below.}, if we add
an additional requirement, namely that the theory should be ``topological''.
This notion may be understood in various ways, cf
e.g.~\cite{Birminghamreview}. Here
we imply the following with it: in addition to the diffeomorphism
invariance of the action functional $L$, we require its moduli space
of classical
solutions (for reasonable enough choice of $\Sigma$, such that
its fundamental group has a finite rank) to be finite dimensional. For
$\dim M > 2$ and generic choice of $\CP$ this is {\em not\/} the case:
For fixed topology of $\Sigma$ the
space of smooth solutions to (\ref{eqs1}) and (\ref{eqs2}) is infinite
dimensional, even after identification of
solutions differing by a gauge symmetry (here only diffeomorphisms of
$\Sigma$).  

The theory described by (\ref{action}) is topological, iff
\cite{PSM0} 
the tensor $\CP$ satisfies \be \CP^{[ij}{},_s \CP^{k]s} = 0 \, ,
\label{Jacobi} \ee i.e.~iff
$\CP$ endows $M$ with a Poisson structure where the Poisson bracket
between functions $f$ and $g$
on $M$ is defined by $\{ f, g \} := f,_i g,_j
\CP^{ij}$. 

Now there are much more gauge symmetries. Indeed, for any choice of
functions $\varepsilon_i(x)$
($\varepsilon$ corresponding to a section of $\CX^* T^*M$), one
verifies easily that if the fields
are transformed infinitesimally according to
\be \delta_\varepsilon X^i = \varepsilon_j \CP^{ji}(X) \; , \; 
\delta_\varepsilon A_i = \md \varepsilon_i + \CP^{kl}{},_{i}(X) A_k
\varepsilon_l
\label{symmetries}\; , \ee
 $L$ changes by a boundary term only. 
The diffeomorphism invariance of $L$ now is a particular
case of (\ref{symmetries}): a simple calculation shows that
on-shell, i.e.~by use of the field
equations (\ref{eqs1}) and (\ref{eqs2}), the Lie derivative of the
fields with respect to  $v \in \Gamma(T\Sigma)$ is  obtained 
upon the (field dependent) choice $\varepsilon_i  = A_{\m i} v^\m$. 

In view of these additional symmetries a new question arises if one
intends to identify (\ref{metric}) with a metric on $\Sigma$: How do
these symmetries act on $g$? If one is ``lucky''---or, more precisely,
upon a reasonable choice of $\CS$---{\em all\/} the symmetries
(\ref{symmetries}), when acting on $g$ obtained as non-degenerate
($\det g \neq 0$ everywhere)
solution to the field equations, will boil down to diffeomorphisms
of $\Sigma$ only. Otherwise there is an additional local symmetry,
which will be hard to interpret in general: There may be e.g.~a flat
spacetime metric $g$ on $\Sigma$ which would be ``gauge equivalent'' to a
spacetime metric $g$ describing some black hole. It is our
intention to clarify the conditions to be placed on $\CS$ which ensure
that this rather unreasonable scenario is avoided.

To be precise:
\begin{defi} \label{defi1}
A Poisson Sigma
Model (PSM) \cite{PSM1,Ikeda}, i.e.~the action functional (\ref{action})
where $\CP$ satisfies (\ref{Jacobi}), together with an identification
(\ref{metric}) for the metric $g$ on $\Sigma$ is called a 
\emph{reasonable theory of 2d Poisson gravity}, if
the following two conditions hold ($d =2$): 
\begin{itemize}
\item For any point $X \in M$ there exist smooth solutions to
the field equations on
$\Sigma \cong \dR^d$ with a base map $\CX \colon \Sigma
\to M$ such that the image of $\CX$ contains $X$ and everywhere on
$\Sigma$ $\det g \neq 0$. 
\item For any two gauge equivalent
solutions  $(\CX_{(1)}, A_{(1)})$ and $(\CX_{(2)},
A_{(2)})$ on a given manifold $\S$ which yield
a non-degenerate metric,  $\det g_{(i)} \neq 0$, 
$\forall x \in \S$, 
and which are maximally extended,
there exists a diffeomorphism $\CD \colon \S \to
\S$ such that $g_{(1)} = \CD^*g_{(2)}$.
\end{itemize}
\end{defi}
We will add  remarks explaining some of the technicalities of this
defintion after its generalization in subsection \ref{sec:defi} below.  
Requiring $(\ref{metric})$ to define a reasonable
theory of gravity will pose some simple geometrical restriction on
$\CS$. This will be seen to define a sub-Riemannian structure
(cf.~e.g.~\cite{Langerock}) on $M$,
which has to be ``compatible'' with the Poisson structure $\CP$ in a
particular manner (equation (\ref{compatibility'}) or (\ref{Poisson}) below). 
{}From a certain perspective to become more transparent below this will
be seen to correspond to a Poisson or Lie algebroid extension of the
notion of a Riemannian foliation (cf.~\cite{Molino}), which may be also
interesting mathematically in its own right.  
Moreover, by the above method we will be able to
define a much wider class (also of just 2d) 
gravity theories, the previoulsy known ones---such as those in
\cite{PSM1,PartI,Habil2,Kummerreview}---arising as a relatively
small subclass.

As a straightforward 
application of this extension it will be immediate to construct a
simple PSM, the gravity
solution space of which coincides
with the one of the exact string black hole, something
excluded within the previously known models according to
\cite{Grumiller:2002md}.
We also demonstrate with an example that by these means
one can construct 
gravitational models with solutions for the metric
that have no local Killing vector field,
a feature missing in the class of matterless 2d gravity models considered
hitherto (cf e.g.~\cite{Habil2,Kummerreview}). We also hope that by
the above analysis much of the in part intricate global structure
arising on $\Sigma$ in the known gravity models, as classified
in \cite{PartII,PartIII}, will
find some explanation from the implicitly underlying geometrical
struture $(M,\CP,\CS)$ on the
target. In fact, already in the present work we want to highlight
the emergence of a Killing vector field in these models as
resulting from structures on $M$.
Finally, we may apply our considerations to the
case of kappa deformed gravity \cite{Mignemi:2002hd,
Mignemi:2002tc}, providing some coordinate
independent \cite{Grumiller:2003df}
grasp on the discussion which arose in that context.

\subsection{Generalizations}
\emph{Given} a metric on $\Sigma$, be it of the form (\ref{metric}) or not,
and the data specified in the previous subsection, we may also add
another term to (\ref{action}): \be \2 \CS^{ij} A_i \wedge \ast A_j \, . 
\label{nontop} \ee Here
$\ast \colon \Omega^p(\Sigma) \to \Omega^{2-p}(\Sigma)$ denotes the
operation of taking the Hodge dual of a form with respect to $g$. In
fact, upon this addition to (\ref{action})
one may obtain a theory that is at least classically equivalent to standard
String Theory: Assuming the matrix $\CT^{ij}$ to be
invertible, we may integrate out the $A$--fields altogether, as they
enter the action quadratically only; the result is the usual String
Theory action in Polyakov form. This may be of particular
interest also if $\CP$ is Poisson, since then one expects $\CP$
to govern the noncommutativity of the effective action induced
on D-branes via the Kontsevich formula \cite{Kontsevich}, which itself
just results upon perturbative quantization of (\ref{action})
\cite{CF1}. For constant $\CT$ this is verified
easily \cite{Volkerinprep}.

We will not consider additions of the form (\ref{nontop}) any further
within the present work. Much closer to the present subject are the
following modifications of the problem posed in the previous
subsection: Suppose that in addition to $\CT$ one is
given also a covariant 2-tensor on $M$. We again may split it
into symmetric and antisymmetric parts, say $\CG$ and $\CB$,
respectively. Then we may add the pullback of $\CB
= \2 \CB_{ij} \md X^i \wedge \md X^j$ with respect to
$\CX \colon \Sigma \to M$ to (\ref{action})
and the pullback of  $\CG = \2 \CG_{ij} \md X^i \md X^j$ to the right
hand side of (\ref{metric}). We then may repeat the questions of the
previous subsection.

Requiring the resulting action functional to be topological now yields
a modified condition (\ref{Jacobi}), its right hand side
consisting of the 3-form $\CH = \md \CB$ with all
indices raised by contraction with $\CP$ \cite{Ctirad}. This leads to
the notion of a twisted Poisson or an $\CH$-Poisson structure on $M$,
cf.~e.g.~\cite{3Poisson}.

Now the second set of field equations receives an $\CH$--dependent
contribution. Likewise, there are still gauge symmetries for any
$\varepsilon_i$, while the transformation on $A_i$ in (\ref{symmetries})
receives some  $\CH$--dependent additions. Despite these evident
changes, much of the standard PSM can be transfered to the modified
theory. Indeed, if $(1 + \CB \CP)$ is invertible, one even may get rid
of the additional term in $L$ by some redefinition of the $A$--fields
(as seen best in the Hamiltonian formulation) leading to a modified,
$\CP$-- and $\CB$--dependent effective Poisson structure $\CP'$. (Such
transformations were called gauge transformations in \cite{3Poisson}).
A likewise statement holds for $g$: Using the first set of field
equations, (\ref{eqs1}), yields a new effective tensor $\CS' = \CS +
\CP \CG \CP$. Thus essentially everything of what will be said on the
PSM with (\ref{metric}) will have a straightforward generalization to
the twisted or HPSM with the modified identification for
$g$.\footnote{In fact, from the mathematical point of view, the more
detailed study of the modified situation still may be  quite
interesting---and maybe also quite intricate (the non-degeneracy
condition on $g$, e.g., will have quite different implications
than in the present note, cf.~Lemma \ref{lemma:nondeg} below).}

Finally, we may also be given a (1,1)--tensor field on $M$. Adding
corresponding terms to (\ref{metric}) does not change much since again
by use of the field equations one may reexpress everything in terms of
a bilinear combination of $A$--fields. If one
uses such a tensor field so as to redefine the first term in
(\ref{action}), on the other hand,
replacing it e.g.~by $A_i \wedge \md X^j \CC^j_i$, the situation
changes if $\CC$ has a kernel. Let us mention here in parenthesis
that such a modification of (\ref{action}) arises if one considers
2d gravity models with additional scalar or fermionic matter fields,
where in the first case there is a nontrivial kernel of $\CC$
considered as a map from $TM$ to $TM$ and in the second case a
nontrivial kernel of the transposed map from $T^*M$ to itself,
cf.~\cite{Pelzer,Habil2}.

More generally we may replace $T^*M$ by some vector bundle $E$ over
$M$. Together with a (coanchor) map $\CC \colon E \to T^*M$ and a
covariant $E$-2-tensor $\CT_{IJ} = \CP_{IJ} + \CS_{IJ}$ we may
replace $(\ref{action})$ by an integral over $A^I \CC_{Ii} \wedge \md
X^i + \2  \CP_{IJ} A^I \wedge A^J$. It may be interesting to clarify
the conditions on such a more general action functional to be
topological and then to address the question of when \be
g = \2 \CS_{IJ} A^I A^J  \label{metric'}
\ee defines a reasonable notion of a metric on $\Sigma$.

Although this generalization of the PSM seems interesting also in its
own right, we will not pursue it here any further. Instead we want
to permit an even more drastic, but different change: We
want to allow for any spacetime dimension of $\Sigma$. Moreover, the
Poisson manifold $(M,\CP)$ may be viewed as a particular Lie algebroid
structure on the vector bundle $T^*M$ over $M$; we will replace it by
any Lie algebroid defined on a vector bundle $E$ over $M$. 

First we note that the field equations (\ref{eqs1}) and (\ref{eqs2})
make perfectly sense also if $\Sigma$ is replaced by a manifold of
some higher dimension, and likewise so for the symmetries
(\ref{symmetries}). So, the questions addressed above do make sense
also in $d$ spacetime dimensions. 

The generalization of the target structure
to Lie algebroids needs some extra explanations; 
before we can approach it, we
want to recall the notion and the basic features of Lie algebroids in the
next section, since we cannot assume that the average physics reader
is familiar with it.

In the present work we will not focus on the
action functional that produces the desired field equations and
symmetries. Whenever an action functional which has field equations and
symmetries \emph{containing} the ones to be specified below,
our analysis will apply. It is still comforting to know, however, that
such action functionals can be constructed \cite{LAYM}.


\subsection{Struture of the paper}
In the following section we will recall the definition of Lie
algebroids over $M$ and provide the corresponding generalization of the
field equations (\ref{eqs1}) and (\ref{eqs2}) as well as of the gauge
symmetries (\ref{symmetries}) to this more general setting. In the
subsequent section we then
determine the conditions on $\CS$ which ensure that $g$ as defined in 
(\ref{metric'}) provides a reasonable theory of gravity,
as defined by the two marked conditions above or more precisely by
Definition \ref{defi1'} below. This section provides the main result
of the present work, summarized in Theorem \ref{theo}.

Section \ref{sec:examples} contains first illustrations of the general
results, applying it to various particular Lie algebroids. Standard
(2+1)-gravity is among the  examples of the general framework
developed here. Only in
Section \ref{sec:Poisson} we come back to the Poisson case---still
permitting arbitrary dimension of $\S$;  in a further
specialization we finally  arrive back to the 
Poisson Sigma model. We then will show how previously known
2d gravity models fit into the present more general framework and provide some
simple examples of new models. Finally we will address
some more recent issues such as kappa deformation of 2d gravity
theories. 
We complete the paper with
an outlook including a list of open questions and possible further
developments.  

\section{Lie algebroids, generalized field equations and symmetries}
\subsection{Lie algebroids, basic facts and examples}
\label{sec:Lie}
A Lie algebroid is a simultaneous generalization of a Lie algebra and
a tangent bundle. Let us begin with a formal definition. (For
further details cf.~e.g.~\cite{SilvaWeinstein}).
\begin{defi}\label{algebroid}
A Lie algebroid $(E,\rho,[ \cdot, \cdot ])$ is a vector bundle $\pi
\colon E \to M$ together with a bundle map (``anchor'')
$\rho \colon E \to TM$ and
a Lie algebra structure $[ \cdot, \cdot ] \colon \Gamma(E)
\times\Gamma(E) \to \Gamma(E)$ satisfying the Leibniz identity
\be
 [ \psi , f \psi'] = f [ \psi, \psi' ] + \left(\rho(\psi) f\right) \psi'
\label{Leibnitz}
\ee
$\forall \psi, \psi' \in  \Gamma(E), \forall f \in
 C^\infty(M)$.
\end{defi}
Let us provide some basic examples of Lie algebroids:
\begin{enumerate}
\item If $M$ is a point, $\rho$ is trivial and
$E$ is just an ordinary Lie algebra.
\item
$M$  arbitrary, but $\rho \equiv 0$, mapping all elements of the
fiber of $E$ at any
point $X \in M$ to the zero vector in $T_XM$, $E$ is a bundle of Lie
algebras, since then (\ref{Leibnitz}) provides a Lie bracket on each
fiber of $E$; $M$ then may be viewed as a manifold of deformations of
a Lie algebra.
\item $E=TM$, $\rho = id_{TM}$, the bracket being the ordinary Lie
bracket between vector fields. This is the so-called standard Lie algebroid. 
\item $E = T^*M$, $(M,\CP)$ a Poisson manifold. Here $\rho = -
\CP^\sharp$, $\rho(\alpha_i \md X^i) = -\alpha_i \CP^{ij}
\partial_j$, and the bracket $[\md f, \md g] := \md \{f,g\}$
between exact 1-forms is extended to all 1-forms by means of
(\ref{Leibnitz}). 
\end{enumerate}
We now add some remarks:
Eq.~(\ref{Leibnitz}) restricts $\rho$; using
the Jacobi identity of the bracket $[ \cdot , \cdot ]$ between
sections, it is possible to show \cite{Boyom}
that $\rho$ is a morphism of Lie algebras,  $\forall \psi, \psi'
\in \Gamma(E)$:
\be \rho [ \psi , \psi' ] =  [ \rho(\psi) , \rho(\psi') ] \, .
\quad  \qquad\label{mor}
\ee
As a consequence, the image of $\rho$ is always an integrable
distribution in $TM$, defining the \emph{orbits} of the Lie algebroid
in $M$. In the case of Poisson manifolds, these orbits coincide with the
symplectic leaves of $(M,\CP)$. 

Generalizing the observation in the second example above, $\ker
\rho \subset E$ always defines a bundle of Lie algebras. For different
points on the same orbit, these Lie algebras are isomorphic, while they
are not necessarily so for any two points in $M$. For Poisson manifolds
$\ker \rho|_{X \in M}$ is the conormal bundle of the respective
symplectic leaf $L$ at a given point $X \in M$. For regular points on
Poisson manifolds these Lie algebras are abelian; the origin of a Lie
Poisson manifold ${\mathfrak g}^\ast$ provides an example for a
nonregular point.  

If $\{ b_I \}_{I=1}^n$ denotes a local basis of $E$, the bracket and
anchor give rise to structure functions $c_{IJ}{}^K(X)$ and $\rho_i^I(X)$,
respectively:
\be [ b_I , b_J ] = c_{IJ}{}^K b_K  , \quad \rho (\partial_i) =
\rho_i^I b_I \, .\ee
The compatibility conditions above then provide differential equations
for them:
\ba \label{compat1}
c_{IJ}{}^S \, c_{KS}{}^L +  c_{IJ}{}^L{},_i \,\rho^i_K \,+ \;
\mbox{cycl}(IJK) &=& 0 \, , \\
c_{IJ}{}^K \rho_K^i - \rho_I^j \r_J^i,_j +  \rho_J^j \r_I^i,_j &=& 0
\, . 
\label{compat2}
\ea
While $\rho^i_I$ behaves as a tensor with respect to a change of local
basis and coordinates, $c_{IJ}{}^K$ certainly does not: With
$b_{\widetilde I} = F^I_{\widetilde I} b_I$ and 
$b_{I} = F_I^{\widetilde I} b_{\widetilde I}$ one has
\be c_{\widetilde K {\widetilde L}}{}^{\widetilde I} = F_{\widetilde K}^K
F_{\widetilde L}^L F^{\widetilde I}_I c_{KL}{}^I + \rho^i_{\widetilde
K} F_{\widetilde L,i}^I F^{\widetilde I}_I - 
\rho^i_{\widetilde L} F_{\widetilde K,i}^I F^{\widetilde I}_I  \, .
\label{trafo}
\ee
In the case
of $\rho \equiv 0$, example 2 above, the second equation becomes
trivial and the first one reduces to the standard Jacobi identity at
any point $X \in M$. For Poisson manifolds, on the other hand, $b_I
\sim \md X^i$, $\rho_J^i \sim \CP^{ji}$, and $c_{IJ}{}^K \sim
\CP^{ij}{},_k$. We see that this case is particular insofar as that the
anchor and the structure functions are essentially one and the same
object; Eq.~(\ref{compat2}) reduces to the Jacobi identity
(\ref{Jacobi}) and  Eq.~(\ref{compat1}) to its derivative.

A Lie algebroid structure on a vector bundle $E$ permits a generalization
of differential geometry to $E$. In particular, there is a natural
exterior $E$-derivative $\md_E\colon \Omega^p_E(M) \to
\Omega^{p+1}_E(M)$, where $\Omega^p_E(M) \equiv \Gamma(\Lambda^p
E^*)$ and  $\Omega^0_E(M) \equiv C^\infty(M)$. If $\{ b^I \}_{I=1}^n$
denotes the local basis of $E^*$ dual to  $\{ b_I \}_{I=1}^n$, it may
be defined by means of
\be \md_E f := f,_i \rho_I^i b^I   \; , \quad \md_E b^I := -\2
c_{JK}{}^I b^J \wedge b^K \; ,
\ee
extended to all of  $\Omega^p_E(M)$ by a \emph{graded} Leibnitz
rule. As a consequence of the Lie algebroid axioms, $\md_E^2 =
0$. ($E$ together with a nilpotent $\md_E$ even provides an alternative
\emph{definition} of a Lie algebroid, cf.~\cite{SilvaWeinstein}).

Likewise there is an $E$-Lie derivative ${}^E\!\CL_\cdot \colon
\Gamma(E) \times \CT^p_q(E) \to \CT^p_q(E)$, where $\CT^p_q(E) \equiv
\Gamma\left(E^{\otimes p} \otimes \left(E^*\right)^{\otimes
q}\right)$. For $p=1, q=0$ it is defined by the ordinary bracket,
${}^E\!\CL_\psi \psi' := [\psi, \psi']$, while for $\a \in
\Omega^p_E(M)$ one uses
the formula (generalizing a well-known pendant in differential
geometry resulting from $E=TM$)
 \be {}^E\!\CL_\psi \alpha = \left(\md_E  \psi\lrcorner +
\psi\lrcorner \md_E\right)\alpha \, ; \ee
it is  extended to  $\CT^p_q(E)$ by
means of the ordinary Leibnitz rule. This concludes our short
excursion into the mathematics of Lie algebroids. 

\subsection{Generalized field equations and symmetries}
For a generalization we first we need to interprete the field content
of the PSM. The coordinates $X^i$ define a map $\CX \colon \S \to M$,
where $\S$ is our spacetime and $M$ the base of a vector bundle
$E$. The fields $A_i$ then correspond to a linear assignment of an
element in $E$ (which is $T^*M$ in the PSM) in the fiber over $\CX(x)$
to any element in $T_x \Sigma$. Together they thus parametrize a
vector bundle morphism
\be 
\varphi \colon T\S \to E \, , \ee
with $\CX$ being the respective base map. Besides $\CX$ such a morphism is
characterized by an element $A=A^I \otimes b_I \in \Gamma(\CX^* E)$,
where $b_I$ here corresponds to a local basis in the pullback bundle
$\CX^* E$ over $\S$ (induced by a likewise basis in $E$). 

As a generalization of the field equations (\ref{eqs1}) and
(\ref{eqs2}), we then consider the following set of equations
\ba
\md X^i - \rho^i_I A^I &=&0 \, , \label{EQS1} \\
\md A^I + \2 c_{JK}{}^I A^J \wedge A^K &=&0 \,  \label{EQS2},
\ea
where $E$ has been given the structure of a Lie algebroid and 
$\rho$ and $c$ are the corresponding previously introduced structure
functions. This clearly generalizes the field equations of the PSM.
Moreover, there is also some mathematical meaning to these equations in the
general case: They state that the vector bundle morphism $\varphi
\colon TM \to E$ is also a morphism of Lie algebroids (cf.~\cite{BKS}
for details).

It is straightforward to check that on behalf of the structural
compatibility equations (\ref{compat1}) and (\ref{compat2}), the above
field equations define a free differential algebra
(cf e.g.~\cite{Sullivan,Izquierdo} for a definition).
In particular, their           
integrability conditions are satisfied and for any choice of
``initial data'' $X^i(x_0)$,
$A^I|_{x_0}$ given at a point $x_0 \in \S$ there is  a neighborhood
$U \ni x_0$ in $\S$ with smooth continuations 
$X^i(x)$, $A^I(x)$ solving the field equations
(\ref{EQS1}) and (\ref{EQS2}).

Such a continuation is however far from unique. There is a set of
symmetry transformations which map solutions of the generalized field
equations (\ref{EQS1}) and (\ref{EQS2}) into solutions. Indeed, one may
check that the infinitesimal transformations
\be
\delta_\varepsilon X^i = \varepsilon^I \rho_I^i \; , \; 
\delta_\varepsilon A^I \approx \md \varepsilon^I + c_{KJ}{}^I A^K 
\varepsilon^J \, ,
\label{Symmetries}\;
\ee
induced by some element $\varepsilon \in \Gamma(\CX^*E)$, leave the
field equations invariant. Here again one needs to use the structural
compatibility conditions, generalizing the Jacobi identity for the
PSM. These are the symmetries we want to consider as gauge
symmetries. 

Note that in the above we used a somewhat modified equality sign
in the definition of the symmetries for the $A$-fields. By this we
intend to imply that the respective symmetry transformations need to
have this form only on-shell, i.e.~upon use of the field equations 
(\ref{EQS1}) and (\ref{EQS2}). Indeed, examples such as the HPSM show
that in general there may be some addition proportional to
(\ref{EQS1}) so as to leave invariant the respective action
functional. Within the present setting it is, however, completely
sufficient to know the gauge symmetries on-shell. 

The above gauge symmetries generate
diffeomorphisms of $\S$. Indeed, again the image of a vector field $v$
under the map $\varphi \colon TM \to E$ provides the respective
infinitesimal generator $\varepsilon$; one finds
\be \CL_v \approx \delta_{\langle A, v \rangle} \; ,
\label{diffeo}
\ee
where $\CL_v$ denotes the standard Lie derivative with respect to $v$,
and $\langle A, v \rangle \equiv A^I_\m v^\m$ the image of $v$ with
respect to $\varphi \sim (\CX, A)$. 

As a further consistency check we may verify that 
the complete set of field equations is covariant. This indeed follows
quite easily by use of Eq.~(\ref{trafo}). Likewise so for the symmetries
(\ref{Symmetries}), keeping in mind their infinitesimal form, however; in
particular,
\be \label{covariant}
\delta (F^{\widetilde I}_I A^I) = 
\delta (F^{\widetilde I}_I) A^I + F^{\widetilde I}_I \delta A^I
 \approx \md \varepsilon^{\widetilde I} + c_{\widetilde K \widetilde L
}{}^{\widetilde I} A^{\widetilde K} \varepsilon^{\widetilde L} \, ,
\ee
where $\varepsilon^{\widetilde I} \equiv F^{\widetilde I}_I
\varepsilon^I$ and $F^{\widetilde I}_I$ is an arbitrary function of
$X(x)$, governing
a change of basis in $E$.  (Although not also of $x$ alone, as it
would be permitted as a change of basis in the pullback bundle $\CX^*
E$---at least if the structure
functions $c_{IJ}{}^K$ in (\ref{Symmetries}) are understood as
pullback of the structure functions on $E$, as we want to interprete
them).

As mentioned already in the Introduction, action functionals
yielding field equations and symmetries containing (\ref{EQS1}),
(\ref{EQS2}), and (\ref{Symmetries}) exist. One option,
which works in any dimension $d$, is to multiply the left hand sides of 
(\ref{EQS1}) and (\ref{EQS2}) by $B_i$ and $B_I$, a $(d-2)$- and a
$(d-1)$-form, respectively (cf also \cite{LAYM} for more details).

\subsection{Definition of a reasonable theory of $E$-gravity} 
\label{sec:defi}
We now are in the position to generalize Definition \ref{defi1}:
\begin{defi} \label{defi1'}
A Lie
algebroid $E$ together with a symmetric section $\CS$ of $\CT_0^2(E)$
is said to define a \emph{reasonable theory of $E$-gravity},
if the two
marked items in Definition   \ref{defi1} hold true for $g$ as defined
in (\ref{metric'}), where solutions refer to the set of
field equations (\ref{EQS1}),(\ref{EQS2}) and the
gauge symmetries are given by (\ref{Symmetries})
(infinitesimally and on-shell).
\end{defi}
We now add some remarks explaining
the two items in the definition: In the first one,
we required the existence of a (metric) non-degenerate
solution in a neighborhood of the preimage of \emph{any}
point $X \in M$. If this were not satisfied, but only for a
submanifold $U \in M$, we could equally well take the subbundle
$E|_U$ as a new Lie algebroid, defined just over $U$. Moreover,
there would be \emph{no} condition on $\CS$ for points
$X \in M \setminus U$. Thus, essentially without loss of generality, we
require non-degeneracy for all points $X \in M$. 
  

Since the second item refers to a global gauge equivalence,
while the symmetries (\ref{Symmetries})
are given in infinitesimal form only, we need to add that
within the present framework and in the context of (\ref{metric'}) we
regard two metric non-degenerate solutions
$(X_{(1)}, A_{(1)})$ and $(X_{(2)}, A_{(2)})$ to the field equations 
as gauge equivalent if they may be connected by a flow of gauge
transformations generated by (\ref{Symmetries}) \emph{which} does not
enter a degenerate sector of the theory, i.e.~which does not pass through a
solution not satisfying the first condition on $g$.

This last specification is necessary because otherwise the set of
theories would be empty: Even if the gauge transformations
$(\ref{Symmetries})$ when acting on \emph{non-degenerate}
solutions correspond merely to diffeomorphisms, they allow to
connect non-degenerate solutions with degenerate ones. E.g.~in the 2d
Poisson case it has been shown that by such
transformations, even if they are contained in the component of 
unity of the gauge group, one may
generate non-degenerate gravity solutions with nontrivial kink number
from non-degenerate ones with trivial kink number
(if $\pi_1(\Sigma) \neq 0$), something that can never be achieved by a
diffeomorphism of $\Sigma$---cf.~\cite{PLB} for an explicit
example of this scenario within the Jackiw-Teitelboim model of 2d
gravity \cite{teitelboim:1983ux,jackiw84},
and \cite{kinks} for a generalization to arbitrary 2d dilaton gravity
theories.  

Finally, the addition that the solution should be maximally extended
results from the following typical feature of a gravity theory:
 the ``time-evolution''---or likewise 
the extension of a solution on $\S$ to one defined on a bigger manifold
$\widetilde \S \supset \S$---is governed by the symmetries, even if
these are on-shell diffeomorphisms only. Thus evolving e.g.~some
region $U \subset \widetilde
\S$ in some locally defined time parameter $t$ such that $U_t
\subset \widetilde \S$ and $U_0 = U$, it may happen that
for some large enough $t=t_1$ one has
$U_1 \cap U = \emptyset$, $U_1 \equiv U_{t_1}$, and that
$g$ on all of $U=U_{0}$ is flat while on $U_1$ it may be nowhere flat.
Then $U_0$ and $U_1$ may be still gauge equivalent with respect to
some symmetries which only on-shell reduce to the diffeomorphism,
while certainly there does not exist a diffeomorphism relating
$(U_0,g|_{U_0}))$ and  $(U_1,g|_{U_1}))$. However, by construction of
this example, there exists some extension $(\widetilde \S, g)$ of
the local solution defined on $\S \cong U_0$ such that
$(U_0,g|_{U_0})$ and
$(U_1,g|_{U_1})$ are part of this extended solution. (Note that,
\emph{as a manifold}, $\widetilde \S$ may still be diffeomorphic
to $\S \cong U_0 \cong U_1$; thus, extending a solution defined on $\S$
need not change its topology). However, if two solutions are already
maximally extendend, then we require that they are gauge equivalent only 
if they (or better their respective metrics) differ by some
diffeomorphism from one another.
Note that by our definition it is not
excluded that two \emph{not} gauge equivalent solutions
still have diffeomorphic
metrics. This can happen, if the model carries also some other
physically interesting fields, such as e.g.~non-abelian gauge
fields. We
do not want to exclude such possibilities.

Alternatively, in the second item we could have taken recourse to
infinitesimal symmetries only. So, we could have required that
for any metric non-degnerate solution $(\CX,A)$, the infinitesimal
variation of $g$, as defined in (\ref{metric'}), by a gauge
transformation (\ref{Symmetries}) may be expressed as the Lie
derivative of some vector field $v \in \Gamma(T\S)$ acting on $g$. 
Despite the technical complications, to our mind the global
description 
is formulated closer to
the desiderata.

Further illustration of the assumptions or requirements in the
definition will be provided by the examples 
below.


\section{Conditions for a reasonable theory of gravity}
\subsection{Metric non-degeneracy}
\label{sec:nondeg}
A metric $g$ on $\S$ is non-degenerate iff $g(v,v')|_x = 0 \; \forall v \in
T_x \S$ implies $v'=0$. Likewise, it is non-degenerate iff the map
$g^\flat \colon T\S \to  T^\ast\S, v \mapsto g(v,\cdot)$ is
invertible. (To be precise, we would have to restrict to the fiber
over a point $x \in T\S$ in the last sentence; this is to be
understood, also analogously in what follows).
Due to (\ref{metric'}), we have $g^\flat = \2 \varphi^*
\circ \CS^\sharp \circ \varphi$, where $\CS^\sharp \colon E \to E^*,
a=a^I b_I \mapsto a^I \CS_{IJ} b^J$ and $\varphi^*$ is the transpose
map to $\varphi$ (again, when restricted to $\CX(x)$, certainly). 
This implies that $\ker \varphi$ has to be the zero vector. Moreover,
we learn that the image of this map is not permitted to have a
nontrivial intersection with the kernel of $\CS^\sharp$. Thus we find:
\begin{lemma} \label{lemma:nondeg}
For (metric) non-degenerate solutions to the field
equations (\ref{EQS1}) and  (\ref{EQS2}), one needs 
\be \ker \varphi = \{ 0 \} \qquad \mbox{and}  \qquad
\mathrm{im} \, \varphi \cap \ker \CS^\sharp = \{ 0 \} \label{cond1}
. \ee
\end{lemma}
As a simple corollary of this, we find that $\dim \mathrm{im} \, \varphi =
d$, where $d$ denotes the dimension of $\S$, and that
$d \le r$, where $r \equiv \mathrm{rk} E$
denotes the rank (fiber dimension) of $E$
and that the
dimension of $\ker \CS^\sharp$ may not exceed the upper bound 
\be  \dim \ker \CS^\sharp \le r - d \quad  \LRA
\quad \dim \mathrm{im} \CS^\sharp \ge d.
\label{dimupper}
\ee
If $\CS$ has definite signature, (\ref{cond1}) is also sufficient to ensure
non-degeneracy of $g$; and then (\ref{dimupper}) is sufficient for the
existence of a non-degenerate solution $\varphi$ at least in some
neighborhood $U \ni x$ of a given point $x \in \S$  due to the local
integrability of the field equations for any given choice of
$\varphi_x \sim (\CX, A)_x$ at that point. Choosing the trivial
topology for $\S$, $\S \approx \dR^d$, we can identify $U$ with $\S$
and then fulfill the first of the two conditions for an $E$-gravity to
be reasonable. 

\begin{prop} \label{prop:ok}
For semi-definite $\CS$ or for any $\CS$ with $\dim \mathrm{im}
\CS^\sharp = d$ 
the conditions (\ref{cond1}) are sufficient to ensure that
(\ref{metric'}) is non-degenerate. 
\end{prop}

In general, however, 
(\ref{cond1}) is not sufficient to ensure non-degeneracy of $g$,
as the following
example may illustrate: Let $d = 2$, $r = 3$, 
$\CS = b_1 b_1 + b_2 b_2 - b_3 b_3$,
such that $\ker \CS^\sharp = \{ 0 \}$, and let
$\mathrm{im} \, \varphi = \langle b_1, b_2+b_3 \rangle$. The conditions
in (\ref{cond1}) are satisfied, but restricting $\CS$ to the
two-dimensional image of
$\varphi$, it reduces to $\CS = b_1 b_1$, which is degenerate.  To
exclude such counter-examples, one may  employ the following
\begin{prop} \label{prop:nondeg} For any $W \subset E$, 
\be \CS |_W \; \textrm{is non-deg.}
\quad \LRA \quad  \CS |_{W \cap Z} \;
\textrm{is non-deg.} \, ,  \ee 
where $Z$ (or better $Z|_{X \in M}$) denotes the set (not a
vector space) of null
vectors, $Z|_X=\{ V \in E_X | \CS(V,V)=0 \}$.   
\end{prop}
\begin{proof}
One direction is trivial, namely $\La$, since for any non-null vector
$V$, one already has $\CS(V,V) \neq 0$. The other direction is an
exercise in elementary linear algebra.  
\end{proof}
So, in addition to (\ref{cond1}), it suffices to check that for any
vector $V \in {\mathrm{im} \,\varphi \cap Z}$, there exists a vector  
$V' \in {\mathrm{im} \,\varphi \cap Z}$ such that $\CS(V,V') \neq 0$ so
as to ensure that the map (or solution) $\varphi\sim (\CX,A)$ leads to
a non-degenerate
metric $g$ (upon usage of the defining relation (\ref{metric'})).

\subsection{Transformation of the metric}
We now come to the derivation of the key relation(s) of the present
paper. Herein we first assume that $\varphi$ is a solution to the field
equations (\ref{EQS1}), (\ref{EQS2}) providing a \emph{non-degenerate}
metric $g$. With the considerations of the previous subsection, it
then will be easy to determine some minimal conditions on $\CS$ such
that this can be achieved at all, which was one of the two conditions
placed on a reasonable theory of $E$-gravity. According to
the second item (cf Defintion \ref{defi1'}), we are then left with
finding conditions on $\CS$ such that the symmetries
(\ref{Symmetries}) coincide with diffeomorphisms of $\S$ on-shell. In
other words, we need to ensure that for any $\varepsilon \in
\Gamma(\CX^* E)$ one either has $\delta_\varepsilon g \approx 0$
\emph{or} there has to exist a $v \in \Gamma(T\S)$ such that
$\delta_\varepsilon g \approx \CL_v g$. According to (\ref{diffeo}) we
may, however, rewrite the right-hand side of the last equation as
another gauge transformation,  $\CL_v g \approx \delta_{\varphi(v)}
g$. Correspondingly, we find that
\be \forall \varepsilon \in
\Gamma(\CX^* E) \quad  \exists \: v \in \Gamma(T\S) \colon  \quad
\delta_{\varepsilon -\varphi(v)}  g
\stackrel{!}{\approx} 0, 
\label{Bed}
\ee
where $v$ may be also the zero vector field. This condition is
sufficient and also necessary so as to ensure that the underlying
assignment for $g$ defines a reasonable theory of $E$-gravity.
We first compute $\delta_\varepsilon g$.
By a straightforward calculation, using the Leibnitz rule for the
symmetries, one obtains
\be \delta_\varepsilon g \approx \2 \CX^*\left({}^E\!\CL_{b_K} \CS \right)_{IJ}
\varepsilon^K A^I A^J +\CX^*(\CS_{IJ}) \md \varepsilon^I A^J \, .
 \label{change}
\ee
If, moreover, $\varepsilon$ is the pull back of a section $\epsilon
\in \Gamma(E)$,
\be \varepsilon^I = \CX^* \epsilon^I  \, , \label{pullback} \ee
then, using the field equations, Eq.~(\ref{change}) may be simplified
further:
\be
 \delta_\varepsilon g \approx \2  \CX^*\left({}^E\!\CL_{\epsilon} \CS
 \right)_{IJ}
 A^I A^J \, . \label{simplified}
\ee 
For a general map $\varphi$ we may \emph{choose} $\varepsilon$ of the
form (\ref{pullback}). On the other hand, for a given map $\varphi$
this covers \emph{any} choice of $\varepsilon$, if the base map $\CX
\colon \S \to M$ is an embedding. Due to the first field equation
(\ref{EQS1}) and due to (\ref{cond1}),
this requires
\be \ker \rho \cap \mathrm{im} \,\varphi = \{ 0 \} \, . \label{reform}
\ee

Let us in the following assume for simplicity that $\CS$ is
semi-definite, i.e.~e.g.~$\CS \ge 0$. (We will soon relax this condition
again; however, much of the discussion simplifies in this case and may
at least serve as a first orientation). Then, due to  (\ref{cond1}), there
is---for fixed, permitted $\varphi$ and for any point $x \in
\Sigma$---a unique orthogonal decomposition of $E_{X(x)}$
according to 
\be E_{X(x)} = W_x \oplus W_x^\perp \, , \quad W_x \equiv \mathrm{im}
\, \varphi|_x \, ,
\label{sum}
 \ee
where a possibly nonvanishing $\ker \CS^\sharp|_{X(x)}$ is part of 
$W_x^\perp$  certainly. Note that this decomposition depends on $x$
and not only on $X(x)$, which may be a decisive difference, if
(\ref{reform}) is not satisfied. Alternatively, we may formulate it as
a decomposition of the pullback bundle:
\be \CX^*E = W \oplus W^\perp \, , \quad W = \mathrm{im}
\, \varphi \, ,
\label{sum'}
 \ee
where here $\mathrm{im}\, \varphi$ is interpreted as subbundle of  $\CX^*E$. 

In the case that $\CX \colon \S \to M$ is an embedding, however,
we may rewrite (\ref{sum}) also in the form 
\be E|_{\mathrm{im}\, \CX} = W \oplus W^\perp \, , \quad W = \mathrm{im}
\, \varphi \, ,
\label{sum''}
 \ee
where now $W$ is a subbundle of $E$ (restricted to the image of $\CX$). 
If not otherwise stated, we will use (\ref{sum'}). 

We know that (\ref{Bed}) has to be
satisfied for any $\varepsilon \colon \S \to E$ covering $\CX \colon
\S \to M$. We now can uniquely decompose any $\varepsilon$ into a part
inside $W$ and $W^\perp$, calling it $\varepsilon_W$  and
$\varepsilon_\perp$, respectively: $\varepsilon =\varepsilon_W +
\varepsilon_\perp$. One has to be careful: if $\varepsilon$ satisfies 
(\ref{pullback}), this in general does \emph{not} imply a likewise
decomposition of $\epsilon$---since the decomposition (\ref{sum})
depends explicitly on $x$ and not only on $X(x)$. This changes,
however, if (\ref{reform}) is satisfied; then also $\epsilon =
\epsilon_W + \epsilon_\perp$. 

It is suggestive to assume that if  (\ref{Bed}) is
to be satisfied, then $\varepsilon_W$ needs to equal
$\varphi(v)$. This is, however, not mandatory: Although there is a
unique decomposition of $\varepsilon$ into the two orthogonal parts,
it does not imply that also the respective variations of $g$ are
orthogonal (so that each of them would need to vanish
separately). Instead we only know that there is some $v'$ such that 
$\varepsilon_W = \varphi(v')$ and then, with $v-v' := w$, we can merely
conclude from (\ref{Bed}) that
\be \delta_{\varepsilon_\perp} g \stackrel{!}{\approx}
\delta_{\varphi(w)} g \quad \mbox{for some} \quad w \in W   \, .
\label{Bed'}
\ee
In the present paper we do not intend to explore this relation in full
generality or even in examples; nor do we want to extend the
general consideration to the case of an indefinite $\CS$, where part of $W$
may be contained in $W^\perp$. Here we will content ourselves with
solving the above condition for $w :=0$, i.e.~we require
\be \delta_{\varepsilon_\perp} g \stackrel{!}{\approx}
0 \qquad  \forall \varepsilon_\perp \in \Gamma(W^\perp) \, . 
\label{Bed''}
\ee
In subsection \ref{sec:main} below, however, we will also permit
indefinite $\CS$, but only for $\dim \ker \CS^\sharp = r-d$. In
\emph{this} case, one may just replace $W^\perp$ by $\ker \CS^\sharp$
in the above. We will make this more explicit later on. 
  
\subsection{First compatibility conditions}
\label{sec:first}
Within this subsection we will first assume
that (\ref{reform}) is satisfied.
Note that this can be achieved only if $\ker \rho \subset E$ is not
too big; in particular, due to the first condition in (\ref{cond1}),
it \emph{cannot} be satisfied if $\dim \ker \rho > r - d$, $r
\equiv \mathrm{rank} E$.

Thus here we may analyse (\ref{Bed''}) for the case of
(\ref{pullback}) with $\varepsilon_\perp = \CX^* \epsilon_\perp$.
Using (\ref{simplified}) in combination with (\ref{Bed''}), we find 
\be 
\left({}^E\!\CL_{\psi_1} \CS \right) (\psi_2, \psi_3) = 0  \quad
\forall \psi_1 \in \Gamma(W^\perp) , \; \psi_2, \psi_3 \in \Gamma(W) 
. \label{compgen} 
\ee
This relation needs to hold for any point $X\in M$, because
any $X$ can be in the image of $\CX$, the base map of $\varphi$. And
according to the above definition of a reasonable theory of
$E$-gravity, by using all possible maps $\varphi$, we may also vary
$W$ essentially at will: Eq.~(\ref{compgen}) has to hold 
\be \forall W \subset E \;
\textrm{with} \;
W \cap \ker \CS^\sharp = \{ 0 \} = W \cap  \ker \rho , \:
\:  \dim W = d  .
\label{forall} \ee

In summary:
\begin{prop}\label{prop:necessary}
In any Lie algebroid $E$ together with a symmetric, semi-definite section
$\CS$ of $E^*\otimes E^*$ defining a reasonable theory of $E$-gravity
(cf.~Definition \ref{defi1'}), the conditions (\ref{dimupper}) and
(\ref{compgen}) with (\ref{forall}) are satisfied. 
\end{prop}
The condition
(\ref{compgen}) with (\ref{forall})
becomes empty certainly
if $\dim \ker \rho > r - d$.
The following observations may be helpful in the present context:
\begin{cor} \label{cor:necessary}
Any semi-definite $\CS$ of a reasonable theory of $E$-gravity with 
$\dim(\ker \rho) \le r -d$ 
satisfies
\be
{}^E\!\CL_\psi \CS = 0 \quad  \forall \psi \in \ker \CS^\sharp \, .
\label{compatibility}
\ee
\end{cor}
\begin{proof}
Take $\psi_1 := \psi \in \ker  \CS^\sharp$ in (\ref{compgen}), which
is possible for \emph{any} choice of $W$, since  always $\ker
\CS^\sharp \in W^\perp$. Next take $\psi_2$ in
(\ref{compgen}) as \emph{any} vector in $E\backslash
\ker \rho \cap E\backslash\ker \CS^\sharp$. Finally, choose $\psi_3$
in such a way that the span of  $\psi_2$ and $\psi_3$ can lie in some
$W$ compatible with (\ref{forall}). Since $d \ge 2$,
$\dim(\ker \rho) \le r -d$, and $\dim(\ker \CS^\sharp) \le r -d$,
the possible choices of $\psi_2$ and
$\psi_3$ lie dense in $E$. By continuity we thus can conclude
$\left({}^E\!\CL_\psi
\CS\right)(\psi_2,\psi_3) = 0$ for all $\psi_2, \psi_3 \in
E$---and thus Eq.~(\ref{compatibility}). 
\end{proof}
\begin{prop} \label{prop:linear}
The conditions (\ref{compgen}) and (\ref{compatibility}) (or
(\ref{compatibility'}) below) are
$C^\infty(M)$-linear. Correspondingly it suffices to check them on a
local basis. 
\end{prop}
\begin{proof} 
In (\ref{compgen}) we only need to check it for $\psi_1$. 
Using the simple-to-verify relation
\be \label{formula}
{}^E\!\CL_{F \psi} \CS = F \;{}^E\!\CL_{\psi} \CS + \md_E F \:
\CS(\psi,\cdot)
\ee
with $\psi=\psi_1$, we find that additional contributions vanish due
to orthogonality of $\psi_1$ with $\psi_2$ and $\psi_3$. 
Likewise, in (\ref{compatibility}) an eventually additional term
vanishes since  $\psi \in \ker \CS^\sharp$.
\end{proof}
A simplification occurs, if  (\ref{dimupper}) is saturated:
\begin{cor} 
 For 
$\dim  \mathrm{im} \CS^\sharp =  d$
 and $\dim(\ker \rho) \le r -d$,
Eq.~(\ref{compgen})
reduces to Eq~(\ref{compatibility}). \label{cor:simple}
\end{cor}
\begin{proof}
In this case always $W^\perp = \ker \CS^\sharp$. Thus (\ref{compgen})
is automatically satisfied upon (\ref{compatibility}), which in turn
was already found to be necessary.  
\end{proof}
The conditions established in the present section are necessary
conditions (for satisfying (\ref{Bed''})) only, which, as remarked,
in the case of 
$\ker \rho$ too big even become vacuous. On the other hand,
when $\ker \rho$ is trivial---so that 
(\ref{reform}) is satisfied always, no matter what $\varphi$---the
conditions in  Proposition \ref{prop:necessary}
are also sufficient to define a reasonable theory of $E$-gravity. 
This is quite restrictive, however: it implies that $E$ is isomorphic to
some integrable distribution, $E \cong D \subset TM$ with
$[\Gamma(D),\G(D)] \subset
\G(D)$. So not many Lie algebroids may be covered by that Proposition.

We will consider the case of injective anchor map in some
detail in the examples. It will be seen by explicit
inspection then that in this case it is \emph{necessary}
that $\ker \CS^\sharp$ has maximal dimension so as to be
compatible with (\ref{dimupper}).
At least in this case it must be
possible, therefore, to derive $\dim  \mathrm{im} \CS^\sharp =  d$
from the conditions summarized in Proposition \ref{prop:necessary}.

Moreover, in the present paper
we were not able to provide examples of admissible $E$-gravity
theories which do not saturate the bound. 
It may be conjectured, therefore, that this is even a necessary
condition in general.
In any case, the conditions
(\ref{compgen}) with (\ref{forall}) seem
very restrictive, and it is plausible that stronger results
can be derived from them.

\subsection{Sufficient conditions}
\label{sec:main}
The main complication one encounters with generalizing the previous
considerations to arbitrary Lie algebroids (also permitting $\dim \ker
\rho > r-d$), and to possibly necessary and sufficient conditions to
ensure (\ref{Bed''}), is the fact that in general the decomposition
(\ref{sum}) depends on $x \in \S$ itself, and not just on its image
$X(x)$ with respect to $\CX \colon \S \to M$. Correspondingly, in
(\ref{change}) it is
not always possible to choose a basis $b_I$ in $E$ adapted to the
decomposition of $\varepsilon$ into its two parts $\varepsilon_W$ and
$\varepsilon_\perp$. (Consider e.g.~the extreme case where the image
of $\CX$ is just a point in $M$). This then can have the effect that
$\CX^*(\CS_{IJ}) \varepsilon_\perp^I A^J = 0$ (which is true by
construction of $\varepsilon_\perp$) does not imply
$\CX^*(\CS_{IJ}) \md \varepsilon_\perp^I A^J = 0$ (the second term in 
Eq.~(\ref{change})). And indeed, as we will also illustrate by an explicit
example in section \ref{sec:bundles} below,
conditions of the form (\ref{compgen}) 
turn out insufficient in general, no matter what $W$ is permitted to
be.

In the following we will thus content ourselves with providing
some sufficient conditions for an $E$-gravity theory to be
reasonable. For this purpose we assume saturation of the general bound
(\ref{dimupper}). This was seen to provide a simplification already in
the previous subsection (cf.~Corollary \ref{cor:simple}). We now state
the main result of the present paper:
\begin{theo}
Any Lie algebroid $E$ together with a symmetric section
$\CS$ of $E^* \otimes E^*$ defines a reasonable theory of $E$-gravity
(as defined in Definition \ref{defi1'}), if $\dim \mathrm{im} S^\sharp
= d$ 
and
\be
{}^E\!\CL_\psi \CS = 0 \quad  \forall \psi \in \ker \CS^\sharp \, .
\label{compatibility'}
\ee
  \label{theo}
\end{theo}
As before, $d = \dim \S$ and $r = \mathrm{rank} \, E$ and
${}^E\!\CL$ 
 denotes the $E$-Lie derivative of the Lie algebroid
as defined at the end of section \ref{sec:Lie}.
\vskip1mm
\begin{proof}[of Theorem \ref{theo}]
We remarked already previously that the symmetries are on-shell
covariant, cf.~Eq.~(\ref{covariant}). Correspondingly, the total
expression (\ref{change}) is independent of a choice of frame $b_I$ in
$E$; the unwanted contribution from the first term cancels a likewise
contribution from the second one on use  of  the field
equations (\ref{EQS1}). Let us therefore choose a basis $b_I$ adapted
to the kernel of $\CS^\sharp \colon E \to E^*$: $\{b_I\}_{I=1}^r = \{ \{
b^{\ker}_A \}_{A=1}^{d-r} , \{ b^{0}_M \}_{M=1}^d \}$, where $ \{
b^{\ker}_A \}$ spans $\ker \CS^\sharp$ and $\{ b^{0}_M \}$ some
$d$-dimensional complement $W^0$. Note that this provides a fixed
decomposition of $E$ according to
\be E = W^0 \oplus \ker \CS^\sharp \, . \label{hilfsdecomp} \ee 
This is in spirit quite different from the decomposition (\ref{sum}),
where for the \emph{same} point $X$ in $M$ there may be different
decompositions into $W_x$ and $W_x^\perp$, depending on the choice of
$x \in \S$. In the present context, always $W_x^\perp=\ker
\CS_{X(x)}^\sharp$, no matter what $x$. On the other hand, in general
$W_x \neq W^0_{X(x)}$. Now we use the unique decomposition of
$\varepsilon$ into its two parts corresponding to  (\ref{sum}).
For the calculation of the change of $g$ with respect to $\varepsilon$
we use the other decomposition (\ref{hilfsdecomp}). We thus need to
express $\varepsilon_W$ and $\varepsilon_\perp$ in terms of the
adapted basis introduced above. While the first of these two
quantities in general has contributions into directions of both $W_0$
and $\ker \CS^\sharp$, which, moreover, depend on $x$ (and not just on
$X(x)$, $\varepsilon_\perp$ appearing in (\ref{Bed''})) has a very
simple decomposition:  $\varepsilon_\perp =
\varepsilon_\perp^A(x) b^{\ker}_A|_{X(x)}$. This has no non-vanishing
components in the $W^0$ direction, and likewise also does not $\md
\varepsilon_\perp^I(x) = (\md
\varepsilon_\perp^A(x), 0)$. Thus obviously (in the adapted basis
chosen) the second term in (\ref{change}) vanishes identically for
$\varepsilon = \varepsilon_\perp$. But
also the first term vanishes on behalf of (\ref{compatibility'}) and
the fact that in the sum over $K$ only contributions along
$b^{\ker}_A$ have non-zero coefficient $\varepsilon_\perp^K(x)$.
This concludes the proof.

(For non-semidefinite $\CS$ one merely replaces  $W_x^\perp$ in 
(\ref{sum}) by $\ker \CS^\sharp_{X(x)}$ so as to guarantee uniqueness
of the decomposition. Cf.~also Proposition \ref{prop:ok} for what concerns
non-degeneracy of $g$.)
\end{proof}
If we restrict our attention to (\ref{Bed''}) (instead of the more
general condition (\ref{Bed'})) then under the assumption on the
dimension of $\ker \CS^\sharp$ the condition  (\ref{compatibility'})
is also \emph{necessary} for defining a reasonable theory of
$E$-gravity. This is obvious from the above proof.

As mentioned in Proposition \ref{prop:linear}, the condition
(\ref{compatibility'}) is $C^\infty$-linear. In fact, this goes even
further\footnote{This observation has been inspired by 
discussions with A.~Weinstein, in particular those
about the standard Lie
algebroid, discussed in section \ref{sec:examples} below, in which he
pointed out to me that in this case the structure induced by a
compatible $\CS$ is the one of a Riemannian foliation.}: 

\begin{prop} \label{prop:Liesub}
For any $\CS \in \Gamma(\vee^2E^*)$ of constant rank
satisfying (\ref{compatibility'}) in a 
Lie algebroid $E$, $K :=\ker \CS^\sharp$ defines a Lie subalgebroid.  
\end{prop}
\begin{proof}
Due to $[{}^E\!\CL_{\psi_1}, {}^E\!\CL_{\psi_2}] =
{}^E\!\CL_{[\psi_1,\psi_2]}$ also the product of two sections
$\psi_1,\psi_2  \in \Gamma(K)$ annihilate $\CS$. Assuming that
$[\psi_1,\psi_2] \not \in\Gamma(K)$, one obtains a
contradiction:  $\exists \psi_3
\in \Gamma(E)$ such that $0 \neq\CS ( {}^E\!\CL_{\psi_1}\psi_2,
\psi_3) = {}^E\!\CL_{\psi_1}
\left( \CS(\psi_2,\psi_3)\right) - 
\CS(\psi_2, {}^E\!\CL_{\psi_1} \psi_3) =0$, since
$\psi_2 \in \ker \CS^\sharp$. 
\end{proof}
As an obvious consequence we find
\begin{cor} $\rho(K) \subset TM$ is an integrable distribution. 
\label{cor:integrable}
\end{cor}

Under appropriate conditions, including that the foliation induced by
$\rho(K)$ is a fibration, this may be used to define quotient
bundles, to which $\CS$ projects as an otherwise arbitrary
nondegenerate $E$-2-tensor. We intend to come back to this elsewhere.

\section{First Examples}
\label{sec:examples}
In this section we provide some elementary examples following the list
after Definition \ref{algebroid}. 

\subsection{Lie algebra case}
\label{sec:Lieal}
In the case of an ordinary Lie algebra, $E = {\mathfrak g}$, only the
second field equation is nontrivial. One then regards flat connections
modulo gauge transformations on a $d$-dimensional spacetime $\S$. In
this way  for $d=3$ one may reobtain standard 2+1 gravity (with or
without cosmological constant) in its
Chern Simons formulation \cite{Achucarro,Witten2+1}, to mention the most
prominent example: If there is no cosmological constant, the
Lie algebra is the one of $ISO(2,1)$, while
in the presence of a cosmological constant $\l$
it is $so(3,1)$ and $so(2,2)$, depending on the sign of $\l$. So, in
each of these cases, the ``rank'' $r$ of the bundle (here just over
a point) is six. The rank of the bilinear form $\CS$ used in
 \cite{Achucarro,Witten2+1} to construct
the metric $g$ on $\S$ is three, on the
other hand, saturating our general bound (\ref{dimupper}). Equation
(\ref{compatibility'}) just reduces to ad-invariance of that
(degenerate) inner product on ${\mathfrak g}$ with respect to those
elements of ${\mathfrak g}$ which are in its kernel ${\mathfrak k}$.
According to Proposition \ref{prop:Liesub}, ${\mathfrak k}$
is a Lie subalgebra of  ${\mathfrak g}$. 

It is worthwhile to  mention that
all of these three Lie algebras
permit an
ad-invariant non-degenerate inner product. This is used to construct
the action functional  \cite{Achucarro,Witten2+1}. However,
this inner product cannot be used to construct the metric $g$---at
least if we want to
comply to the conditions in Theorem \ref{theo}, where we need a
degenerate inner product. For non-zero $\l$, the degenerate bilinear form
 $\CS$ of \cite{Achucarro,Witten2+1} used to construct
the metric is in fact \emph{not} ad-invariant with respect to all of 
${\mathfrak g}$, 
but only for 
${\mathfrak k} < {\mathfrak g}$.

\subsection{Bundles of Lie algebras}
\label{sec:bundles}
This is the case where $E$ is some vector bundle over $M$, equipped
with a smoothly varying Lie algebra on the fiber. The field equations
(\ref{EQS1}) and (\ref{EQS2}) then just imply that $\S$ has to be
mapped to one (arbitrary) point $X_0$ in $M$ and that the $A$--field
is a flat connection of the Lie algebra in the fiber above
$X_0$. Gauge transformations are again just of the standard Yang-Mills
type, where the point $X_0$ cannot be changed (although there is a
solution for any point $X_0 \in M$). The type of the Lie algebra
(although not its dimension, since this is fixed by the rank of the
bundle $E$) can change upon a different choice of  $X_0$. 

Let us further specialize to the case of abelian Lie algebras,
i.e.~all the fibers carry the same, trivial Lie algebra
$[ \cdot , \cdot ] \equiv 0$. Then, irrespective of the choice of
$X_0$,  one finds as local solutions for the
$A$-fields, $A^I = \md f^I$, $i=1, \ldots , r$, for some functions
$f^I$. Gauge transformations correspond to $A^I \sim A^I + \md g^I$
for some arbitrary functions $g^I$.

We want to work out this example in full detail, making a general
ansatz for $\CS$, \be \CS = \2 \CS^{IJ}(X) b_I b_J \, , \label{S}
\ee and then determine
the conditions to be fulfilled such that the respective $E$-gravity
becomes reasonable  (cf.~Definition \ref{defi1'}). This then will be
compared with the general results obtained in the previous
section. Note that 
in the present case $\ker \rho = E$, so that we
cannot refer to the (necessary) conditions obtained in
subsection \ref{sec:first}; we can only refer to \ref{sec:main},
which, however, had the drawback that it provided only sufficient
conditions---due to the restriction on the rank of $\CS$. In the
present example we will find this restriction to be also necessary.
In the case of bundles of abelian Lie algebras, moreover, the condition
(\ref{compgen}) (or also (\ref{compatibility'})) is \emph{empty},
since the $E$-Lie derivative vanishes identically.

Let us now work out this example explicitly. With (\ref{metric'}) we find
that any solution for $g$ is of the form
\be g = \2 \CS_{IJ}(X_0) \, \md f^I \md f^J \label{explicit}
\ee
in the present case.  First we require the
existence of non-degenerate solutions. For this  $\CS(X_0)$ needs
to have at least rank $d\equiv \dim \S$---and in the
spirit of the remark
at the end of section \ref{sec:nondeg}---for all $X_0 \in M$. This
coincides with what we found in (\ref{dimupper}). (For simplicity we
take $\CS$ to have positive semi-definite signature so that this
condition is also sufficient.)

To simplify the further discussion, let us in the following
diagonalize the matrix $\CS(X_0)_{IJ}$: This can be done most easily
by transition to new functions representing the solutions for $A$: 
$f^I(x) \to  M^I_J(X_0) f^J(x)$. The net effect in (\ref{explicit}) is
that $\CS(X_0) = \mathrm{diag}(1,\ldots,1,0, \ldots , 0)$, where $m\ge
d$ entries are non-vanishing. Now a 
possible non-degenerate representative is obtained on $\S \approx
\dR^d$ upon the choice $f^I =  x^J$ for $I = 1,
\ldots, d$ and $f^I=0$ for $I > d$; then \be g = \2 dx^\m
dx^\m, \label{flat} \ee which is just the standard flat metric on $\dR^d$.


Next we need to check the behavior of $g$ under gauge
transformations. We at once observe that with respect to gauge
transformations (\ref{Symmetries}), on a topologically trivial $\S$,
any $g \sim 0$. A likewise statement holds also for 2+1 gravity in its
Chern Simons formulation discussed above or 2d dilaton gravity in its
Poisson Sigma formulation recalled below---and we thus do not want to
exclude such a feature.
\emph{However}, restricting the gauge transformations to the
non-degenerate sector, i.e.~disregarding all solutions to the field
equations which somewhere on $\S$ have $\det g =0$, we want the gauge
transformations to boil down to diffeomorphisms of $\S$, and to nothing
else. 

How or under what conditions is this realized in the present
simple example? Indeed, suppose
first that the rank $m$ of $\CS$ equals $d$. Then any permitted $g$ is
of the form  $g= \2 \sum_{\m=1}^d \md f^\mu \md f^\mu$ with
$\det g \neq 0$, which
is equivalent to $\md f^1 \wedge \ldots \wedge \md f^r \neq 0$ in this
case. Clearly, such a $g$ is always flat, and \emph{any} two such
non-degenerate solutions are related to one another by a
diffeomorphism of $\S$. 

Now let us assume that $m > d$. We clearly obtain the above
flat solution $g$ always, too, since we can just require all $f^I$
with $I > d$ to vanish. But then any other non-degenerate solution
$g$ on $\dR^d$ should
be \emph{diffeomorphic} to this flat solution (since the moduli space of
solutions with respect to the symmetries (\ref{Symmetries}) consists
of one point here and in a reasonable theory of gravity these two
notions coincide---up to the subtleties mentioned), i.e.~be flat too.
This, however, is not the case for  $m > r$: A straightforward
calculation shows that the curvature of $g_{\m\n} = \delta_{\m\n}  
 + f,_\m f,_\n$ is non-zero (for sufficiently general second
derivatives of $f$). This metric, however, is a gauge relative to
(\ref{flat}) with respect to the symmetries (\ref{Symmetries}),
provided only $m>r$ (and if $f$ is kept sufficiently small, it never
passes through a non-degenerate sector).

In summary we find in this simple case:
\begin{prop} For $E$ being a bundle of abelian Lie algebras,
$\dim \mathrm{im} \CS^\sharp = d$ is necessary and sufficient 
for defining an admissible  $E$-gravity theory. 
\end{prop}


Let us use the occasion to also illustrate the possible complications arising
from the $x$-dependence of the decomposition (\ref{sum}).  Let us
choose $\S = \dR^2$, $E=M\times \dR^3$, and $\CS_{IJ} = \delta_{IJ}$.
Consider $A^1 = \md x^1$, $A^2 = \md x^2$, $A^3 = \md f$ with $f \in
C^\infty(\dR^2)$. Thus $W_x = \langle \partial_1 + f,_1 (x)
\partial_3, \partial_2 + f,_2 (x)
\partial_3 \rangle$. Clearly, the following gauge parameter is in
$W^\perp_x$: $\varepsilon =\varepsilon_\perp = (f,_1 , f,_2,-1)$ (the
entries corresponding to the standard basis $\{\partial_i\}_{i=1}^3$
in the fiber of $E$). We now turn to (\ref{change}): 
the first term in (\ref{change}) vanishes identically, whatever
$\varepsilon$. In this example, the second term is already on-shell
covariant by itself (due to $\md X^i \approx 0$). By construction,
$\CX^* \CS_{IJ} \varepsilon_\perp^I A^J \approx 0$. However,
e.g.~$\delta_{\varepsilon_\perp} g_{11} \approx \delta_{IJ}
(\varepsilon_\perp^I),_1 A^J_1 = f,_{11}(x)  \not \approx 0$ (for
nonvanishing second derivatives of $f$ with respect to $x^1$). This
may be contrasted with the proof of Theorem \ref{theo}, where the
second term in (\ref{change}) was found to vanish (upon an appropriate
choice of a basis) no matter how
complicated the $x$-dependence of the decomposition (\ref{sum}).

In the present example the
resulting solutions for $g$ were found to be  flat always---for
admissible theories. This will
change in general, if the Lie algeras are taken non-abelian. 

\subsection{The standard Lie algebroid and integrable distributions}
\label{sec:standard}
We first consider the illuminating case $E=TM$. 
The field equations (\ref{EQS1})  reduce to $\md X^i =
A^i$ in this case, while the second set of field
equations, (\ref{EQS2}), becomes an identity.
Thus, in the present case, using (\ref{metric'}) we find
\be \label{pull}
g \approx \2 \CS_{ij} \md X^i \md X^j = \CX^* \CS \, , \ee
$g$ is just the pullback of the section $\CS \in \Gamma(\vee^2 T^*M)$
to $\S$ by the map $\CX$.

Let us consider first $n=d$, $n\equiv \dim M$. Then by condition
(\ref{dimupper}), $\CS$ is non-degenerate and defines a metric on
$M$ and $g$ is nothing but the pullback of this metric to $\S$.
Note that this provides a simple example where the necessity for
the addition of maximal extension of $\S$ in the second item of
Definition \ref{defi1} or \ref{defi1'} becomes quite transparent
(cf.~also the discussion at the end of
subsection \ref{sec:defi}): $\CS$ may be quite different for
different regions in $M$. 
After maximal extension, $\CX \colon \S \to M$ is a (possibly branched)
covering map (thus locally a diffeomorphism). If one does not permit the
branched coverings (e.g.~by requiring solutions to be either
geodesically complete or that curvature invariants diverge towards the
ideal boundary---assuming that this is satisfied for $\CS$ on $M$), so
as to exclude constructions of solutions such as those in \cite{kinks}, the
moduli space of classical solutions in this example is zero
dimensional, independetly of the topology of
$\S$. It may consist, however, of several representative
solutions, parametrized by the homotopy (covering) classes of the map $\CX$.


We now turn to $d<n$. It is obvious that for 
 $\dim \mathrm{im}
\CS^\sharp > d$ the conditions of a reasonable theory of
$E$-gravity cannot be satisfied, since then we can embed some
maximally extended $\S$ in various ways such that $\CX^* \CS$ will
give non-equivalent (i.e.~non-diffeomorphic) metrics on
$\S$. Correspondingly, we find upon explicit inspection that in the
case of $TM$ we need saturation of 
the bound
$(\ref{dimupper})$.

The condition (\ref{compatibility}) just becomes an ordinary Lie
derivative on $M$ in the present context. $\CS$ has to be invariant
with respect to motions in directions of its kernel. This kernel is
integrable according to Corollary \ref{cor:integrable}. $\CS$ then
provides a metric in the normal bundle of the leaves of  the
respective foliation. This metric is invariant along the foliation,
moreover; it thus projects to locally defined quotient manifolds (we
are dealing with regular foliations, since the rank of $\CS$ is
constant; thus \emph{locally} the foliation is a fibration and a quotient
manifold, the base of the fibration, can be defined), on which it is
nondegenerate, moreover. Such a structure on $M$
is called a Riemannian foliation, cf.~e.g.~\cite{Molino}
(or then probably pseudo-Riemannian foliation if 
 the locally projected metric has indefinite signature).

Let us provide a
simple example of such a feature on $M=\dR^3$ coordinatized by 
$(X,Y,Z)$: $\CS := \md X^2 + \md Y^2$. Clearly this is invariant with
respect to $\partial_Z$, $\CL_Z \CS = 0$, which spans its kernel;
according to Proposition \ref{prop:linear} and Corollary
\ref{cor:simple} this is already sufficient
for a check. The foliation in this simple example is a fibration over
$\dR^2 \ni (X,Y)$ with fiber $\dR \ni Z$. Thus the quotient manifold
here exists even globally, and it is equipped with the non-degenerate
metric $\md X^2 + \md Y^2$ (the projection of $\CS$). 

In this example we assumed $d=2$, certainly. The
representative metric $g$ on $\S$ is the unique flat metric in this
case  (if it
exists); whatever embedding of $\S$, for a non-degenerate solution one
needs $\md X \neq 0$ and $\md Y \neq 0$, and then $X$ and $Y$ can be
taken as possible coordinates on $\S$. Note that in this example the
topology of $\S$ permitting everywhere non-degenerate solutions for
$g$ is quite restricted; in fact, there is e.g.~no compact Riemann
surface for $\S$ with this property. Note that at least locally we can
identify a solution on $\S$ with the quotient manifold constructed
above. Different embeddings of $\S$ (provided homotopic along the
$Z$-fibers) are just gauge equivalent; and this is also obvious from
the explicit solutions since in any solution as constructed above $g$
is independent of the function $Z$.

In the present context one may
certainly also easily obtain non-flat metrics $g$ on $\S$; e.g., just
multiply the above $\CS$ by some non-trivial function $f(X,Y)$.

As the standard Lie algebroid $E=TM$ in general is one of the main
models for a Lie algebroid, the guideline for it besides ordinary
Lie algebras, it may be used also in the present context for further
orientation in the general Lie algebroid case. Equipping a Lie algebroid $E$
with an $E$-tensor $\CS$ satisfying the conditions in
Theorem \ref{theo} might thus be called equipping $E$ with
an $E$-Riemannian foliation.

Qualitatively, the situation does not change much when we regard 
integrable distributions $D \subset TM$: First we observe that
according to (\ref{EQS1}) $\CX$ maps $\S$ completely into a
leaf $L$ of the distribution. The moduli space of classical solutions
will now consist of homotopy classes of such maps. But otherwise the
discussion is essentially as before, just---for any fixed base map
$\CX$---with $TM$ replaced by $D|_L \cong TL$. Note that now $\CS$ is
not just a covariant 2-tensor on $M$. Rather, it is the equivalence
class of such 2-tensors, where two 2-tensors are to be identified, if
they coincide upon contraction with arbitrary vectors tangent to the
distribution $D$. As for an explicit example we may extend the
previous one for $TM$ by adding one or more further directions to $M$,
considering the previous $M$ as a typical leaf $L$. 

\begin{prop} For Lie algebroids with injective anchor map $\rho$ the
conditions in Theorem \ref{theo} are also necessary.
\label{prop:injective}
\end{prop}

\section{Specializing to Poisson gravity}
\label{sec:Poisson}
\subsection{Compatibility and Integrability}
We first want to specialize (\ref{compatibility'}) to the Poisson
case. Here $E=T^*M$, $M$ the given Poisson manifold, and $\CS$ becomes
a symmetric, contravariant $2$-tensor on $M$,  $\CS = \2 \CS^{ij}
\partial_i \partial_j$. Correspondingly, $\CS^\sharp \colon T^*M \to
TM$ and  $\mathrm{im} \CS^\sharp \subset TM$ defines a distribution on
$M$. In contrast to the distrubtion of $\CP^\sharp$, which induces the
symplectic foliation of $M$, this distribution is not necessarily
integrable. To provide a non-integrable example, just consider $d=2$, 
$M=\dR^3$ with the trivial Poisson bracket---so that
the condition (\ref{compatibility'}) becomes empty---and take $\CS =
(\partial_1 + X^3 \partial_2) \partial_3$.

The structure defined on
$M$ by $\CS$ (with constant rank) is called a sub-Riemannian (or in
the indefinite case then sub-pseudo-Riemannian)  structure,
cf.~e.g.~\cite{Langerock}. According to Theorem \ref{theo} it has to
satisfy a particular compatibility condition with the Poisson
structure on $M$. We will now make this explicit under a further
assumption, namely  that $\mathrm{im}
\CS^\sharp$ is an integrable distribution. (In the context of
sub-Riemannian structures this is usually assumed to \emph{not} be the
case; still, here we assume this for simplicity.)
Then at least locally one
may characterize the corresponding leaves, which we want to denote by
$R$, by the level set of some
functions $f^\a \in C^\infty(M)$, $\a = 1 \ldots n-d$, where $n \equiv
\dim M$. (Likewise the symplectic leaves will be denoted by $L$ or
$(L,\O_L)$, where $\O_L$ is the symplectic 2-form induced by the
Poisson tensor $\CP$ on $M$; locally any leaf $L$ can be characterized
as the level set of some Casimir functions $C^I$, $I = 1 \ldots n - k$
where $k = \mathrm{rank} \CP_X$ for any point $X\in L$.)

For any symmetric or antisymmetric (say contravariant) 2-tensor $\CT$,
the annihilator of
its image $\mathrm{im} \CT^\sharp \subset TM$ is 
its kernel, $\ker \CT^\sharp \subset T^*M$. In the above case we thus
can span the kernel of $\CS^\sharp$ by $\md f^\a$, $\ker \CS^\sharp 
= \langle \md f^\a \rangle_{\a=1}^{n-d}$. The quotient map $\CS^\sharp
\colon T^*M/\langle \md f^\a \rangle_{\a=1}^{n-d} \to \mathrm{im}
\CS^\sharp$ has an inverse. Since at any point $T^*R$ may be
identified with $\colon T^*M/\langle \md f^\a \rangle_{\a=1}^{n-d}$,
this implies that the leaves $R$ are equipped with some
pseudo-Riemannian metric. Let us denote the latter by ${\cal G}_R$.

So, in this scenario, $M$ is foliated in two different ways, by
symplectic leaves, $M = \bigcup (L,\O_L)$, and by Riemannian ones, 
$M = \bigcup (R,{\cal G}_R)$. While the former foliation may be quite
wild, the leaves of the latter ones all have at least the same
dimension, namely $d\equiv \dim \Sigma$.
(We will content ourselves here
with specializing the conditions of Theorem \ref{theo} to the Poisson
case---under the additional assumption of integrable
$\mathrm{im} \CS^\sharp$). The situation simplifies further if
the pseudo-Riemannian foliation is even a fibration, such that
in particular the quotient manifold $M/R$ is well-defined (and leaves 
$R$ lying dense in some higher dimensional submanifolds of $M$ are
excluded). 

The two foliations (or, more precisely, the
two structures induced on $M$ by $\CP$ and $\CS$, respectively)
are not independent from one another, due to
condition (\ref{compatibility'}):
\begin{prop}
Let $(M,\CP)$ be a Poisson manifold and $\CS \in \Gamma{\left(
\vee^2 TM\right)}$
define a $d$-dimensional fibration into pseudo-Riemannian leaves
$(R,{\cal G}_R)$ as described above. Then $(M,\CP,\CS)$ define a
reasonable theory of Poisson gravity, if
\be \CL_{v{_f}} \CS  = 0 \qquad \forall f \in C^\infty(M/R),
\label{Poisson}
\ee
where $v_f \equiv \{ \cdot , f \}  = \CP^{ij} f,_j \partial_i$ is the
Hamiltonian vector field of the function $f$ and $\CL$ denotes the
usual Lie derivative.   \label{prop:Poisson}
\end{prop}
To prove this statement it suffices to check that for \emph{closed}
1-forms $\a \in \O^1(M)$ one has ${}^E\!\CL_\a = 
\CL_{\rho(\a)}$---since
$\rho(\md f) = v_f$.\footnote{I am grateful to A.~Kotov for discussions of
this point. Cf.~also Eq.~(\ref{Poissonlie}) below.}
Due to Proposition \ref{prop:linear} it is
sufficient to check (\ref{Poisson}) for the set of $n-d$ 
functions $f^\a$ introduced above. 

Let us add two remarks, valid in the simplified circumstances
of the above Proposition. First, 
according to Proposition \ref{prop:Liesub}, $\ker S^\sharp=
\langle \md f^\a \rangle_{\a=1}^{n-d}$ defines a
Lie subalgebroid. It is isomorphic to the Poisson type Lie algebroid
$T^*(M/R)$, where, by construction, the projection $\pi \colon M \to
M/R$ is a Poisson map (since $f^\a$ can be thought of as pullback of
at least locally defined functions on $M/R$---and since $\ker \CS^\sharp$
is a Lie subalgebroid of $T^*M$) \footnote{I am grateful to
A.~Weinstein for this remark.}. Second, according to Corollary
\ref{cor:integrable}, $\CP^\sharp (\ker \CS^\sharp)$
defines an integrable distribution; it is spanned by the vector fields
$v_f$ appearing in Eq.~(\ref{Poisson}). In general this foliation is 
distinct from the symplectic and from the Riemannian foliation. Clearly, its
leaves always lie in symplectic leaves $L$, however. Its relation to
the Riemannian leaves is given by
\begin{prop} The leaves of $\CP^\sharp (\ker \CS^\sharp)$ lie inside
the intersection $L \cap R$ of symplectic and Riemannian leaves, iff
$R$  are coisotropic submanifolds of $(M,\CP)$.  \label{coiso}
\end{prop}
\begin{proof} $v_{f^\a} \in \mathrm{im} \CS^\sharp  \LRA  \langle
\ker  \CS^\sharp , v_{f^\a} \rangle = 0   \LRA  \{f^\a , f^\b \} =
0$. 
\end{proof}
As a rather obvious consequence of this we find 
\begin{prop} The (pseudo-)Riemannian structures on the leaves $R$ may
be chosen
independently from one another  (still varying smoothly certainly)
iff the leaves $R$ are
coisotropic submanifolds of the Poisson manifold $M$. In this case the
metric on each leaf $R$ has $k$ independent Killing vector fields,
where $k =  \dim \CP^\sharp (\ker \CS^\sharp)$. \label{prop:Killing}
\end{prop}
Note that in the present Poisson case this does \emph{not} imply
automatically that also the metric $g$, defined on $\S$, has Killing
vector fields. In fact, by the field equations, $\S$ is mapped into
the symplectic leaves, which in general are different from the
Riemannian ones; moreover, $g$ is not just an ordinary
pullback. Cf.~also Section \ref{sec:Killing} below for illustration. 

In Proposition \ref{prop:Poisson} we assumed a fibration. More generally, for
a foliation one has a likewise statement, if one uses for any  $X \in
M$  sufficiently small neighborhoods $U_X$ and replaces the above
functions $f$ by functions constant along connected components of
$R \cap U_X$ for any leaf $R$.

We conclude this subsection with an obvious remark:
\begin{cor} \label{cor:triv}
If each leaf of the foliation induced by $\CS$ 
consists of a union of symplectic leaves (the leaves induced
by $\CP$), the condition
(\ref{compatibility'}) or (\ref{Poisson}) is fulfilled trivially.
\end{cor}
This just corresponds to the particular case $k=0$ in Proposition
\ref{prop:Killing}. 

\subsection{Examples in two dimensions}
\label{sec:2d}
Most of the studied
2d gravity models result from the following simple choice of
$(M,\CP,\CS)$ \cite{PSM1}: Take $M = \dR^3$ with linear coordinates
$(X^a,X^3)$, where the index $a$ runs over two values, which
conventionally we denote by 1 and 2 when we consider Euclidean
signature gravity and by $+$ and $-$ for Lorentzian signature; these
indices will be raised and lowered by means of the standard
flat Riemannian or Lorentzian metric $\eta_{ab}$, furthermore. The
Poisson structure is then defined by 
$\{X^a, X^b\} = \varepsilon^{ab} \, W(X^cX_c,X^3)$ 
and
\be \{X^a, X^3\} = \varepsilon^{ab} X_b
 \, , \label{Lorentz} \ee
where $\varepsilon^{ab}$ are the (contravariant) components of the
antisymmetric $\varepsilon$-tensor and $W$ is any smooth two-argument
function. With an ansatz $\CS := \l^{ab}(X) \partial_a \partial_b$,
$\ker \CS^\sharp = \langle \md X^3 \rangle$. Eq.~(\ref{Lorentz}) shows
that $X^3$ generates (Lorentzian or Euclidean) rotations in the planes
of constant $X^3$. According to  Proposition \ref{prop:Poisson}, $\CS$
has to be invariant with respect to these ``rotations''. As one may
easily convince oneself, requiring $\CS$ to be smooth on \emph{all} of
$\dR^3$ leaves only
\be \CS = \2 \g(X^cX_c,X^3) \, \eta^{ab} \partial_a \partial_b \label{2d} \ee
for some nonvanishing two-argument function $\g$. In most applications 
(cf.~below) either $\g=1$ or $\g$ depends on $X^3$ only.

In this example, the symplectic foliation is given by variation of the
integration constant in the solution of the first order differential
equation $\md u/\md v = W(u,v)$, replacing $u$ by $X^aX_a$ and $v$ by
$X^3$ \cite{PartI}. E.g.~for $W=V(X^3)$ one finds $X^aX_a -
\int^{X^3} V(v) \md v = const.$ as for the symplectic leaves. They are
generically two-dimensional, except at simulatenos zeros of $X^a$ and
$W(0,X^3)$, where the Poisson tensor vanishes altogether. On the
two-dimensional leaves the symplectic form can be written as $\O_L=\md
X^3 \wedge \md \varphi$ where in the Euclidean case $\varphi$ is the
standard azimuthal angular variable around the $X^3$ axis and in the
Lorentzian case e.g.~$\ln X^+$. 

The pseudo-Riemannian leaves are characterized by $X^3=const$ in the
above ansatz and $\CG_R = \frac{1}{2\g} \eta_{ab} \md X^a \md
X^b$. Note that the intersection of the symplectic and the Riemannian
leaves are rotationally invariant. So, e.g.~for Euclidean signature
they are generically circles in the $X^3$-planes around the origin
$X^a=0$. 

\subsection{Relation to 2d Lagrangians}
\label{sec:2dL}
We want to be rather brief here due to the rather extensive
literature on this subject (cf.~e.g.~\cite{Kummerreview} and
\cite{Habil2} for two recent reviews), deferring more details also to a
separate work \cite{5peopleinprep}. Still it is illustrative to show
how intricate the relation to some more established actions may be.

First of all we just mention that geometrical action functionals of
the form $L[g] = \int_\S \md^2x \sqrt{|\det g|} f(R)$ or also
\be L[g,\tau^a] \sim L[e^a,\o] = \int_\S
\md^2x \sqrt{|\det g|} \, F(R,\tau^a \tau_a) \label{tor}
\ee may be covered by the
choice in the previous subsection with $\gamma \equiv 1$. 
Here $R$ denotes the Ricci scalar (in two dimensions this contains
all information about the curvature tensor), in the first case
of the torsion-free Levi-Civita connection of $g$ and in the
second case of the connection $\o$ with torsion scalar $\tau^a$.
The functions $f$ and $F$ are the Legendre transforms of the functions
$V$ and $W$ of the preceding subsection, respectively. In these
examples one identifies $A_a$ with a zweibein $e_a$ and $A_3$ with the
single non-trivial component $\o$ of the spin connection, using the
action functional (\ref{action}) Poisson Sigma Model. We refer to
\cite{Habil2} for a detailed study of this relation, including a
discussion of the situation when the functions $f$ or $F$ have Legendre
transforms locally only. 

Secondly, also dilaton-like Lagrangians
\be \label{dilaton} L[g,\phi]= \int_\S \md^2x \sqrt{|\det g|} \left[
\phi R + W((\nabla \phi)^2,\phi)  
\right] 
\ee
can be covered. At least if $W$ in
(\ref{dilaton}) is at most linear in its first argument, there are various
ways of doing so. In one of these, one usually performs a
dilaton-dependent conformal transformation of $g$ and then identifies
$A_i$ in (\ref{action}) with the zweibein and spin connection of the
new metric ) and $X^3$ with the dilaton $\phi$
(while $X^a$ become Lagrange multipliers for torsion zero;
in this case
$W_{\textrm{Poisson}}(X^aX_a,X^3) = W_{\textrm{above}}(0,X^3)$,
cf.~e.g.~\cite{PartI} for further details). 
Closer to the present point of view is to regard this as a
one-step-procedure, using the conformal factor as an $X^3$-dependent
$\gamma$ in (\ref{2d}). 

There is also another route to (\ref{dilaton}), which moreover works
for general $W$ (which now agrees with the function introduced
in the Poisson structure within the previous subsection),
cf.~e.g.~\cite{Kummerreview}. We believe that it is worthwhile to be a
bit more explicit in this case. We first choose to identify $g$ with
\be  g = \2 \eta^{ab} A_a A_b \, , \label{g2d} \ee
corresponding to $\g=1$ in (\ref{2d}). Then let $\omega(A)$ denote the
solution of $\md A^a + \varepsilon^{ab} \omega \wedge A_b = 0$---such
a solution always exists and is even unique (interpreting $A_a$ as
zweibein, this is the corresponding torsion-free connection
1-form). Next, shift $A_3$ by $\omega(A)$ in (\ref{action}), $A_3 \to
\widetilde A_3 := A_3 + \omega(A)$. Then, with the previous choice of
$\CP$, the action functional takes the form (again we replace $X^3$ by
$\phi$) 
\ba L[X,A] &=& \int_\S \omega(A) \wedge \md \phi +  \2  \varepsilon^{ab}
A_a \wedge A_b
 \, W(X^cX_c,\phi)  \nonumber \\ && +
\widetilde A_3 \wedge \left(
\md \phi + \varepsilon_{ab} X^a A^b \right)\, . \label{new}\ea
Classically it is always permitted to eliminate fields of an action
functional by their \emph{own} field equations. We apply this to
$\widetilde A_3$ and $X^a$: variation with respect to $\widetilde A_3$
determines $X^a$ uniquely, and vice versa. Implementing this into
(\ref{new}), the last term vanishes and $X^a$ is replaced by
$-\varepsilon^{a}{}_b A^b_\m g^{\m\n} \partial_\n \phi$. After a
partial integration in the first term, this is now seen to coincide
with (\ref{dilaton}). Note that the elimination of $X^a$ required that
$A^a_\m$ has an inverse (this is reflected by the use of $g^{\m\n}$ in
the above explicit expresion), which is legitimate after one identifies $g$
as in (\ref{g2d}); the equivalence of action functionals then
certainly may be expected also only on the metric non-degenerate
solutions. 

If in the first term of (\ref{dilaton}) $\phi$ is replaced by some
function $U(\phi)$, one still may use the functional (\ref{action})
along the above lines, where now $X^3 = U(\phi)$. Certainly this works
only in regions of $\phi$ where $U$ is monotonic. If $U$ has several
monotonic parts, one needs to use several action functionals to
recover all the classical solutions. However, at least if
$W$ in  (\ref{dilaton})
is at most linear in its first argument,
$W=V_1(\phi) + V_2(\phi)(\nabla \phi)^2$, it may be shown \cite{Norbert}
 under rather mild conditions on $U$, $V_1$, and $V_2$, that for any
 solution with $\phi$ taking values within one of the monotonic
 sectors of $U$, the classically allowed values of $\phi$ remain in
 this sector. 

\subsection{Further 2d examples}
\label{sec:further}
Up to now, 2d models where considered merely with the 3d Poisson manifold
$(M_3,\CP^{(W)})$ as provided in
subsection \ref{sec:2d}---with only one exception, namely
the inclusion of Yang-Mills fields \cite{PartI}. In this case the
total Poisson manifold has the form $(M,\CP) =M_3 \times {\mathfrak g}^\ast$,
where the function $W$ is permitted to depend also on the Casimir
functions of the Lie Poisson manifold ${\mathfrak g}^\ast$. If
$\gamma$ in (\ref{2d}) depends on $X^aX_a$, $X^3$ and again these
Casimirs only, then also the condition (\ref{Poisson}) in Proposition
\ref{prop:Poisson} is satisfied and $(M,\CP,\CS)$ defines a reasonable
theory of 2d Poisson gravity.

But certainly many more examples or models can be constructed by the
general methods of the present paper, even in the realm of two
dimensions. Let us use this fact to construct a simple PSM, the
local solution space of which coincides with \emph{any} given
parametric family of 2d metrics, say $g(C_\a) =  \2
g_{\m\n}(C_\a,x^1,x^2)
\,\md x^\m \md x^\n$ in some local coordinate system $x^\m$ and with 
$C_\a$, $\a = 1 \ldots N$ local parameters in some given
$N$-dimensional manifold $\cal{M}$ of moduli. Just take $M = {\cal{M}}
\times \dR^2$, equipping the former factor with zero Poisson bracket
and $\dR^2$ with the standard one: $(q,p) \in \dR^2$,  $\{ q,p \} =1$,
$\{ q,q \} = 0 = \{ p,p \}$. Now choose $\CS$ as
$$ \CS = \2 g_{11} \partial_q \partial_q + 
\2 g_{22} \partial_p \partial_p - 
 g_{12} \partial_q \partial_p\, , $$ 
where all the coefficient functions $g_{\m\n}$ are evaluated at $(C_\a,
q,p)$. Due to Corollary \ref{cor:triv} this provides a reasonable
theory of 2d gravity. With the simple field equations $\md q + A_p = 0$, 
$\md p- A_q = 0$, $dC^\a = 0$, following from (\ref{eqs1})---and the
fact that locally the second set of field equations (\ref{eqs2}) may
be satisfied without the addition of any further moduli parameters
when using the symmetries (\ref{symmetries})---one obtains the desired
result.

In \cite{Grumiller:2002md} it was shown that for (\ref{dilaton}) 
one finds the (local) solution space 
of the exact string black hole for no choice of $W$; a model which
does the job is
trivially included in our framework, as just demonstrated for an
arbitrary family of 2d metrics. (This can be easily
adapted to also include the given parametric dependence of the dilaton
field $\phi$). Likewise constructions also work in any dimension $d$,
moreover. 

In \cite{Mignemi:2002hd,Mignemi:2002tc}, on the other hand, the
Poisson bracket on $\dR^3$ recalled above was modified, in particular
breaking (or ``$\kappa$-deforming'') its rotational invariance. This
is used hand in hand with the unmodified notion
(\ref{g2d}) for the metric. Correspondingly, the conditions for a
theory to be admissible or ``reasonable'' in our sense and as made
precise in Defintion \ref{defi1'} are violated---with the
corresponding features, too. 

To provide an even simpler example of what happens when these
conditions are violated, consider e.g.~$\dR^3 \ni (q,p,C)$ with
$\CP = \2\partial_q \wedge \partial_p$ and $\CS = \partial_q
\partial_p + f(q,p) \partial_C \partial_C$ for some fixed,
non-constant function $f$. Here $\ker \CS^\sharp$ is trivial, and
$\ker \rho = \langle \md C \rangle$. Now the necessary conditions of
Proposition \ref{prop:necessary} are violated: Indeed, take $W=\langle
\md q, \md p + \md C \rangle$ such that $W^\perp = \langle
\md (p - C) \rangle$; now (\ref{compgen}) is violated due to  
$\left(\CL_{\partial_q} \CS\right)( \md p + \md C,  \md p + \md C)
\propto \partial_q f \neq 0$. The general local solution for $g = \2
\CS^{ij} A_iA_j$, on the other hand, takes the form $g = \md q^2 + \md
p^2 + f \md \l^2$, where $\l$ is an arbitrary function of $q$ and
$p$ and pure gauge according to (\ref{symmetries}). Clearly, for $\l
\equiv 0$ the metric is flat, while otherwise it is generically not.
(Note that this example of a non-reasonable theory works also for
$f=1$, while then the necessary conditions found in Proposition
\ref{prop:necessary} are too weak to exclude this case). 

As we want to stress in the present note, the action functional
(\ref{action}) of the PSM alone certainly does not entail any notion
of a metric $g$ on $\S$; and the identification of $g$ implicitly
introduces or requires further structures. These structures are
identified as a symmetric contravariant 2-tensor on $M$ in the present
paper. Moreover, the 2-tensor may not be chosen at will, but has to
satisfy a compatibility condition with the chosen Poisson structure.
The present framework is independent of coordinates on $M$,
thus also permitting a coordinate independent discussion of
compatibility (cf.~\cite{Grumiller:2003df}). 

\subsection{Relation of Killing vectors to target geometry}
\label{sec:Killing} 
It has been observed \cite{PartI} that all solutions of
\emph{any} gravity model resulting from the structures
as defined in subsection \ref{sec:2d}---and thus,
as a consequence, of the action functionals
(\ref{tor}) and (\ref{dilaton})---have
at least one local Killing vector field $v \in \G(T\S)$.
Moreover, along these lines, $X^3$ is constant; correspondingly, the
Killing lines are apparently related to the one-dimensional
intersection of the symplectic with
the pseudo-Riemannian leaves. (Recall that classically only
maps from $\S$ into a symplectic leaf is compatible with (\ref{eqs1}),
while the pseudo-Riemannian leaves are characterized by $X^3 =
const$). This is qualitatively different from the construction of models
in subsection \ref{sec:further} for any given family of
metrics $g$ on $\S$, in particular also metrics without
Killing symmetry: there the symplectic foliation agreed with the
pseudo-Riemannian one and the intersection was two-dimensional.
It is the purpose of the present subsection to
relate the appearance of a Killing vector field in the models of
subsection \ref{sec:2dL} to the target geometry
$(M,\CP,\CS)$ defined in subsection \ref{sec:2d}. 

We will make this relation explicit in the somewhat simplified
scenario of Eq.~(\ref{reform}), which, on behalf of the field
equations (\ref{eqs1}), implies that \emph{locally} we may
identify $\S$ with a symplectic leaf $L$ and we likewise may
fall back on (\ref{pullback})  and (\ref{simplified}). In the present
case $\ker \CS^\sharp=\langle \md X^3 \rangle$. Let us denote a
local Casimir function of the symplectic foliation by $C$, then
$\ker \rho = \ker \CP^\sharp = \langle \md C \rangle$. Now, for any
point $X \in \textrm{im}\CX \subset M$ there
is a \emph{unique} $\l(X)$ such that $(\l \md X^3 + \md
C)_{X(x)} \in \textrm{im} \varphi_x \subset T_{X(x)}^*M$. Thus,
for any $\m \in C^\infty(M)$ we have $\s := \m (\l \md X^3 + \md
C) \in  \textrm{im} \varphi$. Let us calculate the variation of
$g$ with respect to $\CX^* \s$:
\be \delta_{\CX^* \s} g \approx  \2  \CX^*\left({}^E\!\CL_{\s} \CS
 \right)^{ij}  A_i A_j =  \2  \CX^*\left({}^E\!\CL_{\m \md C} \CS
 \right)^{ij}
 A_i A_j, \label{Killing}
 \ee
where we made use of Eq.~(\ref{compatibility'}). We know that the left
hand side may be rewritten as the Lie derivative of a (for fixed $\s$)
unique (local) vector field $v$ on $\S$, since $\s$ was in the image
of $\textrm{im} \varphi$ and we then may employ (\ref{diffeo}). Thus,
we will have accomplished our task of
showing the existence of a local Killing vector field provided that
the right hand side of Eq.~(\ref{Killing}) can be made to vanish. Note
that in this process the ambiguity in $\m \in C^\infty(M)$ has to be
cut down since in the space of Killing vector fields is a finite
dimensional vector space (and not a module over the functions).
For any $\a \in T^*M$, one has 
\be {}^E\!\CL_{\a} \CS =  \CL_{\rho{(\a)}} \CS + \CP^{is} (\md
\a)_{st} \CS^{tj} \partial_i \partial_j \, .  \label{Poissonlie}
\ee
Using $\a = \m \md C$ and the fact that this $\a$ is in the kernel of
$\rho$, we obviously only need to ensure that $\md \a = \md \m \wedge
\md C = 0$ which is satisfied, iff $\m = \m(C)$. Since on the
classical solutions $C=const$, this corresponds only to a constant
rescaling of the Killing vector.

This concludes our proof for the case of (\ref{reform}). To show the
appearance of this Killing vector field more generally, one 
proceeds similarly to the proof of Theorem \ref{theo}, replacing  
(\ref{hilfsdecomp}) by $E = \md C \oplus \bar W^0 \oplus \md X^3$
(which may be used at least locally on $M$ and for $\md C \not \propto
\md X^3$).

\section{Conclusion}
In the present paper we discussed conditions which can be used to
endow a large class of topological models with some gravitational
interpretation. Concerning the models we only specified their field
equations and local symmetries in the present paper, so as to remain
as general as possible (while still action functionals can be
constructed for them, at least if one permits auxiliary
fields). They could be assigned to any $d$-dimensional
spacetime manifold $\S$ and to any Lie algebroid $E$ used as target. 
On some more abstract level the field equations state that the maps
$\varphi \colon T\S \to E$ should be Lie algebroid (co)morphisms
(i.e.~the induced map $\Phi \colon  \Omega^p_E(M) 
\to \Omega^p(M)$ should be a chain map), while the
symmetries correspond to a notion of homotopy of such morphisms
(cf.~\cite{BKS} for more details).

Our ansatz for the metric $g$ on $\S$ was that it should be the image
of a symmetric, covariant $E$-2-tensor $\CS$,
 $\CS \in \Gamma(\vee^2 E^*)$,  with respect to
the naturally extended map $\Phi$ just introduced above. ($g =
\Phi(\CS)$; in explicit formulas this gives (\ref{metric'})).

We attempted to clearly formulate the desiderata of what we
want to call a reasonable theory of $E$-gravity
(defined over $\S$)---cf.~Defintion \ref{defi1'}. Essentially it was
the requirement that on solutions the gauge transformations of the
model when applied to a non-degenerate metric $g$
should boil down to just diffeomorphisms of $\S$. 

In Theorem \ref{theo} we summarized our main result: If $\CS$ has a
kernel of maximal rank so as to be still compatible with metric
non-degeneracy (cf.~Eq.~(\ref{dimupper})), 
\be \dim \mathrm{im} \CS^\sharp = d  \label{satur},  \ee
then it suffices to have $\CS$ invariant with respect to any section
in its kernel, cf.~Eq.~(\ref{compatibility'}). This last condition was
found also to be necessary (to be precise, cf.~Corollary
\ref{cor:necessary}, this was shown only
under the assumption $\dim \ker \rho \le r-d$ and
by requiring the somewhat strengthened condition (\ref{Bed''})). 
We believe that Eq.~(\ref{satur}) is necessary, too; however, in the
present paper, we managed to show this  only under the  relatively
strong additional assumption that $\ker \rho$ is trivial (cf.~Proposition
\ref{prop:injective}) or that it is all of $E$ and the bracket abelian
(cf.~section \ref{sec:bundles}). 

The invariance condition (\ref{compatibility'}) may be reformulated
as a kind of ad-invariance of the degenerate inner product
induced by $\CS$: Let $(\psi_1,\psi_2) := \langle \CS , \psi_1 \otimes
\psi_2 \rangle$, then Eq.~(\ref{compatibility'}) is equivalent to
\be \rho{(\psi)} \cdot (\psi_1,\psi_2) = ( [\psi , \psi_1],\psi_2) + 
(\psi_1,[\psi,\psi_2]) \, \label{adinv} \ee
valid $\forall \psi \in \Gamma(\ker \CS^\sharp)$
 and for all sections $\psi_1$,  $\psi_2$; here $\cdot$ denotes
application of the vector field to the respective function to follow.  

It is decisive that this condition has to hold only for the Lie
subalgbroid $\ker \CS^\sharp$; if e.g.~it were to hold for any section of
$E$ and if the inner product were non-degenerate, then
\emph{necessarily} $\rho \equiv 0$, i.e.~one would be left with
a bundle of Lie algebras equipped with a fiberwise ad-invariant inner
product. Moreover, as illustrated in section \ref{sec:Lieal} and
\ref{sec:bundles}, even for $\rho \equiv 0$ a kernel of $\CS$ is/will
be required (for $r > d$) so as to really yield a reasonable
(cf.~Definition \ref{defi1'})
notion of a metric $g$ on $\S$ by means of $g = \Phi(\CS)$.

General ad-invariance of an inner product  appears in a slight
modification of the notion
of a Lie algebroid, namely the Courant algebroid; 
modifying the bracket $[ \cdot, \cdot
]$  by a symmetric  contribution stemming from the inner product, one
can have ad-invariance even without $\rho \equiv 0$ and despite
non-degeneracy of the inner product, cf.~e.g.~\cite{Roytenbergdiss,3Poisson}. 
Let us remark in parenthesis that from a more recent perspective,
cf.~\cite{AS}, it is also natural to
drop the non-degeneracy condition; Lie algebroids then appear 
as completely degenerate Courant algebroids. In any case, it may be
reasonable to generalize the considerations of the present paper to
the realm of Courant algebroids. 


Closer to the present structure are however possible quotient
constructions. The guideline may be taken from the standard Lie
algebroid $E=TM$. Assuming $\ker \CS^\sharp$ to define a fibration,
then $\CS$ obviously is just the pullback (by the projection)
of a \emph{non-degenerate 
and unrestricted} metric on the base or quotient manifold. This
generalizes: As one may show and as shall be made more explicit
elsewhere, one may construct a reasonable theory of $E$-gravity
just by means of a Lie algebroid morphism from $E$ to some Lie
algebroid $\bar E$ of rank $d$, where the latter is equipped with an
\emph{arbitrary} non-degenerate fiber metric.  

For $E=TM$ the structure governing a gravity model was found to be the
one of a Riemannian foliation (cf.~section \ref{sec:standard}). The
general structure
required then may be viewed as the corresponding Lie algebroid
generalization, i.e.~an $E$-Riemannian foliation. In the case of
$E=TM$, the metric $g$ was found to locally coincide just with the
metric on the locally defined quotient of the Riemannian foliation.
In general, the relation between $\CS$ and $g$ is more subtle, and
worth further investigations. 


The usually
considered gravity models in three spacetime dimensions result from
$E$ being a Lie algebra (cf.~section \ref{sec:Lieal}). 

For $E$ the cotangent bundle of a Poisson manifold, finally, a
reasonable choice of $\CS$ boilt down to the choice of a
sub-Riemannian structure, compatible with the given Poisson
structure. Under
the assumption that $\mathrm{im} \CS^\sharp$ defines an integrable
distribution, a sub-(pseudo)-Riemannian structure gives rise to a 
foliation of $M$ by $d$-dimensional (pseudo)-Riemannian leaves.
(This is not to be confused with the foliation generated by $\rho \ker
\CS^\sharp$, cf.~the discussion after Proposition \ref{prop:Poisson}.)
The
invariance condition (\ref{adinv}) then was tantamount to requiring
that $\CS$---or the
degenerate inner product induced by it---is invariant with respect to the
Hamiltonian vector fields of the (possibly only locally defined) functions
characterizing the pseudo-Riemannian foliation (cf.~\ref{prop:Poisson}). 

Known models of 2d gravity were seen to be a special case of the much
more general construction of the present paper. Also we were able to
relate the existence of a Killing vector of $g$ in the previously
known models to the invariance condition (\ref{adinv}) together with
the particularities of the two foliations defining these models. In
view of Proposition \ref{prop:Killing}, it may be rewarding to strive
at a generalization of these results. 

It is worthwhile also to generalize our consideration to the
existence of a vielbein $e^a$ and a spin connection
$\o^a{}_b$---instead of just a metric $g$, which results from the
vielbein according to $g=e^a e^b \eta_{ab}$,
i.e.~to determine the
conditions such that some of the 1-form fields $A^I$ permit
identification with the Einstein-Cartan variables $(e^a,
\o^a{}_b)$. 

The present paper is meant to open a new arena for studying models of
gravity. It should be possible to extend the in part extensive studies
of lower dimensional models of gravity---on the classical and
on the quantum level---to a much larger set of
theories, including topological gravity models in arbitrary spacetime
dimensions. 

Finally, we mention again that within this paper we focused on the
field equations and symmetries corresponding to 
Lie algebroids. Depending on the spacetime dimension and possibly
additional structures chosen, there may be several different
action functionals producing the same 
desired theory, at least in a subsector.

To give a simple example:
If $E$ is a Lie algebra ${\mathfrak g}$, the field equations
correspond just to flat
connections and the symmetries to the standard non-abelian gauge
transformations. This may be described without any further structure
and restriction to spacetime by a $BF$-theory. For $d=3$ and with
${\mathfrak g}$ permitting a non-degenerate, ad-invariant inner
product, one may also take a Chern-Simons action; this even has the
advantage of getting by without additional auxiliary fields---but on
the other hand requires additional structures and restriction to
a specific spacetime dimension.

This generalizes also to the present 
context. We intend
to make this more explicit elsewhere \cite{LAYM}.

Merely on the level of the field equations and symmetries, one may already
address the interesting but also challenging
question of observables. Since the theories constructed
are topological, they are expected to capture important and
interesting mathematical information--displaying an interplay between
the topology of $\S$ and the target Lie algebroid and, if one also
involves $g$, the target $E$-Riemannian
structure. As suggested by the particular case $E={\mathfrak g}$, one
should, among others, introduce a generalization of the notion of a Wilson
loop for this purpose.

\begin{acknowledgments}
First of all I need to acknowledge the profit I have drawn from a
collaboration with Martin Bojowald and Alexei Kotov on closely related
subjects, which we are going to publish seperately \cite{BKS}.
In addition, I am grateful for discussions with 
D.~Grumiller, M.~Gr\"utzmann, 
W.~Kummer, D.~L\"ange,
S.~Mignemi, D.~Roytenberg, H.~Urbantke,
D.~Vassilevich and in particular with A.~Weinstein.

\end{acknowledgments}

\bibliography{bibfile}

\begin{thebibliography}{37}
\expandafter\ifx\csname natexlab\endcsname\relax\def\natexlab#1{#1}\fi
\expandafter\ifx\csname bibnamefont\endcsname\relax
  \def\bibnamefont#1{#1}\fi
\expandafter\ifx\csname bibfnamefont\endcsname\relax
  \def\bibfnamefont#1{#1}\fi
\expandafter\ifx\csname citenamefont\endcsname\relax
  \def\citenamefont#1{#1}\fi
\expandafter\ifx\csname url\endcsname\relax
  \def\url#1{\texttt{#1}}\fi
\expandafter\ifx\csname urlprefix\endcsname\relax\def\urlprefix{URL }\fi
\providecommand{\bibinfo}[2]{#2}
\providecommand{\eprint}[2][]{\url{#2}}

\bibitem[{\citenamefont{Birmingham et~al.}(1991)\citenamefont{Birmingham, Blau,
  Rakowski, and Thompson}}]{Birminghamreview}
\bibinfo{author}{\bibfnamefont{D.}~\bibnamefont{Birmingham}},
  \bibinfo{author}{\bibfnamefont{M.}~\bibnamefont{Blau}},
  \bibinfo{author}{\bibfnamefont{M.}~\bibnamefont{Rakowski}}, \bibnamefont{and}
  \bibinfo{author}{\bibfnamefont{G.}~\bibnamefont{Thompson}},
  \bibinfo{journal}{Phys. Rept.} \textbf{\bibinfo{volume}{209}},
  \bibinfo{pages}{129} (\bibinfo{year}{1991}).

\bibitem[{\citenamefont{Schaller and Strobl}(1994{\natexlab{a}})}]{PSM0}
\bibinfo{author}{\bibfnamefont{P.}~\bibnamefont{Schaller}} \bibnamefont{and}
  \bibinfo{author}{\bibfnamefont{T.}~\bibnamefont{Strobl}}
  (\bibinfo{year}{1994}{\natexlab{a}}),
  \eprint[http://arXiv.org/abs]{gr-qc/9406027}.

\bibitem[{\citenamefont{Schaller and Strobl}(1994{\natexlab{b}})}]{PSM1}
\bibinfo{author}{\bibfnamefont{P.}~\bibnamefont{Schaller}} \bibnamefont{and}
  \bibinfo{author}{\bibfnamefont{T.}~\bibnamefont{Strobl}},
  \bibinfo{journal}{Mod. Phys. Lett.} \textbf{\bibinfo{volume}{A9}},
  \bibinfo{pages}{3129} (\bibinfo{year}{1994}{\natexlab{b}}),
  \eprint{arXiv:hep-th/9405110}.

\bibitem[{\citenamefont{Ikeda}(1994)}]{Ikeda}
\bibinfo{author}{\bibfnamefont{N.}~\bibnamefont{Ikeda}}, \bibinfo{journal}{Ann.
  Phys.} \textbf{\bibinfo{volume}{235}}, \bibinfo{pages}{435}
  (\bibinfo{year}{1994}), \eprint{arXiv:hep-th/9312059}.

\bibitem[{\citenamefont{Langerock}(2002)}]{Langerock}
\bibinfo{author}{\bibfnamefont{B.}~\bibnamefont{Langerock}}
  (\bibinfo{year}{2002}), \eprint[http://arXiv.org/abs]{math.DG/0210004}.

\bibitem[{\citenamefont{Molino}(1988)}]{Molino}
\bibinfo{author}{\bibfnamefont{P.}~\bibnamefont{Molino}},
  \emph{\bibinfo{title}{Riemannian {F}oliations}}
  (\bibinfo{publisher}{Birkh\"auser}, \bibinfo{address}{Basel},
  \bibinfo{year}{1988}).

\bibitem[{\citenamefont{Kl{\"o}sch and Strobl}(1996{\natexlab{a}})}]{PartI}
\bibinfo{author}{\bibfnamefont{T.}~\bibnamefont{Kl{\"o}sch}} \bibnamefont{and}
  \bibinfo{author}{\bibfnamefont{T.}~\bibnamefont{Strobl}},
  \bibinfo{journal}{Class. Quant. Grav.} \textbf{\bibinfo{volume}{13}},
  \bibinfo{pages}{965} (\bibinfo{year}{1996}{\natexlab{a}}),
  \eprint{arXiv:gr-qc/9508020}.

\bibitem[{\citenamefont{Strobl}(1999)}]{Habil2}
\bibinfo{author}{\bibfnamefont{T.}~\bibnamefont{Strobl}}
  (\bibinfo{year}{1999}), \bibinfo{note}{{H}abilitationsschrift, 224 pages},
  \eprint[http://arXiv.org/abs]{hep-th/0011240}.

\bibitem[{\citenamefont{Grumiller et~al.}(2002)\citenamefont{Grumiller, Kummer,
  and Vassilevich}}]{Kummerreview}
\bibinfo{author}{\bibfnamefont{D.}~\bibnamefont{Grumiller}},
  \bibinfo{author}{\bibfnamefont{W.}~\bibnamefont{Kummer}}, \bibnamefont{and}
  \bibinfo{author}{\bibfnamefont{D.~V.} \bibnamefont{Vassilevich}}
  (\bibinfo{year}{2002}), \eprint[http://arXiv.org/abs]{hep-th/0204253}.

\bibitem[{\citenamefont{Grumiller and Vassilevich}(2002)}]{Grumiller:2002md}
\bibinfo{author}{\bibfnamefont{D.}~\bibnamefont{Grumiller}} \bibnamefont{and}
  \bibinfo{author}{\bibfnamefont{D.}~\bibnamefont{Vassilevich}}
  (\bibinfo{year}{2002}), \eprint[http://arXiv.org/abs]{hep-th/0210060}.

\bibitem[{\citenamefont{Kl{\"o}sch and Strobl}(1996{\natexlab{b}})}]{PartII}
\bibinfo{author}{\bibfnamefont{T.}~\bibnamefont{Kl{\"o}sch}} \bibnamefont{and}
  \bibinfo{author}{\bibfnamefont{T.}~\bibnamefont{Strobl}},
  \bibinfo{journal}{Class. Quant. Grav.} \textbf{\bibinfo{volume}{13}},
  \bibinfo{pages}{2395} (\bibinfo{year}{1996}{\natexlab{b}}),
  \eprint{arXiv:gr-qc/9511081}.

\bibitem[{\citenamefont{Kl{\"o}sch and Strobl}(1997)}]{PartIII}
\bibinfo{author}{\bibfnamefont{T.}~\bibnamefont{Kl{\"o}sch}} \bibnamefont{and}
  \bibinfo{author}{\bibfnamefont{T.}~\bibnamefont{Strobl}},
  \bibinfo{journal}{Class. Quant. Grav.} \textbf{\bibinfo{volume}{14}},
  \bibinfo{pages}{1689} (\bibinfo{year}{1997}).

\bibitem[{\citenamefont{Mignemi}(2002{\natexlab{a}})}]{Mignemi:2002hd}
\bibinfo{author}{\bibfnamefont{S.}~\bibnamefont{Mignemi}}
  (\bibinfo{year}{2002}{\natexlab{a}}), \eprint{hep-th/0208062}.

\bibitem[{\citenamefont{Mignemi}(2002{\natexlab{b}})}]{Mignemi:2002tc}
\bibinfo{author}{\bibfnamefont{S.}~\bibnamefont{Mignemi}}
  (\bibinfo{year}{2002}{\natexlab{b}}), \eprint{hep-th/0210213}.

\bibitem[{\citenamefont{Grumiller
  et~al.}(2003{\natexlab{a}})\citenamefont{Grumiller, Kummer, and
  Vassilevich}}]{Grumiller:2003df}
\bibinfo{author}{\bibfnamefont{D.}~\bibnamefont{Grumiller}},
  \bibinfo{author}{\bibfnamefont{W.}~\bibnamefont{Kummer}}, \bibnamefont{and}
  \bibinfo{author}{\bibfnamefont{D.~V.} \bibnamefont{Vassilevich}}
  (\bibinfo{year}{2003}{\natexlab{a}}), \eprint{hep-th/0301061}.

\bibitem[{\citenamefont{Kontsevich}(1997)}]{Kontsevich}
\bibinfo{author}{\bibfnamefont{M.}~\bibnamefont{Kontsevich}}
  (\bibinfo{year}{1997}), \eprint[http://arXiv.org/abs]{q-alg/9709040}.

\bibitem[{\citenamefont{Cattaneo and Felder}(2000)}]{CF1}
\bibinfo{author}{\bibfnamefont{A.~S.} \bibnamefont{Cattaneo}} \bibnamefont{and}
  \bibinfo{author}{\bibfnamefont{G.}~\bibnamefont{Felder}},
  \bibinfo{journal}{Commun. Math. Phys.} \textbf{\bibinfo{volume}{212}},
  \bibinfo{pages}{591} (\bibinfo{year}{2000}), \eprint{arXiv:math.qa/9902090}.

\bibitem[{\citenamefont{Schomerus and Strobl}(1999)}]{Volkerinprep}
\bibinfo{author}{\bibfnamefont{V.}~\bibnamefont{Schomerus}} \bibnamefont{and}
  \bibinfo{author}{\bibfnamefont{T.}~\bibnamefont{Strobl}}
  (\bibinfo{year}{1999}), \bibinfo{note}{unpublished notes}.

\bibitem[{\citenamefont{Klimcik and Strobl}(2002)}]{Ctirad}
\bibinfo{author}{\bibfnamefont{C.}~\bibnamefont{Klimcik}} \bibnamefont{and}
  \bibinfo{author}{\bibfnamefont{T.}~\bibnamefont{Strobl}},
  \bibinfo{journal}{J. Geom. Phys.} \textbf{\bibinfo{volume}{43}},
  \bibinfo{pages}{341} (\bibinfo{year}{2002}), \eprint{math.sg/0104189}.

\bibitem[{\citenamefont{Severa and Weinstein}(2001)}]{3Poisson}
\bibinfo{author}{\bibfnamefont{P.}~\bibnamefont{Severa}} \bibnamefont{and}
  \bibinfo{author}{\bibfnamefont{A.}~\bibnamefont{Weinstein}},
  \bibinfo{journal}{Prog. Theor. Phys. Suppl.} \textbf{\bibinfo{volume}{144}},
  \bibinfo{pages}{145} (\bibinfo{year}{2001}), \eprint{math.sg/0107133}.

\bibitem[{\citenamefont{Pelzer and Strobl}(1998)}]{Pelzer}
\bibinfo{author}{\bibfnamefont{H.}~\bibnamefont{Pelzer}} \bibnamefont{and}
  \bibinfo{author}{\bibfnamefont{T.}~\bibnamefont{Strobl}},
  \bibinfo{journal}{Class. Quant. Grav.} \textbf{\bibinfo{volume}{15}},
  \bibinfo{pages}{3803} (\bibinfo{year}{1998}), \eprint{gr-qc/9805059}.

\bibitem[{\citenamefont{Strobl}(2003)}]{LAYM}
\bibinfo{author}{\bibfnamefont{T.}~\bibnamefont{Strobl}}
  (\bibinfo{year}{2003}), \bibinfo{note}{in preparation}.

\bibitem[{\citenamefont{da~Silva and Weinstein}(1999)}]{SilvaWeinstein}
\bibinfo{author}{\bibfnamefont{A.~C.} \bibnamefont{da~Silva}} \bibnamefont{and}
  \bibinfo{author}{\bibfnamefont{A.}~\bibnamefont{Weinstein}},
  \emph{\bibinfo{title}{{G}eometric {M}odels for {N}oncommutative {A}lgebras}},
  vol.~\bibinfo{volume}{10} of \emph{\bibinfo{series}{Berkeley Mathematics
  Lecture Notes}} (\bibinfo{publisher}{American Mathematical Society,
  Providence, RI}, \bibinfo{year}{1999}), \bibinfo{note}{available at
  http://www.math.berkeley.edu/~alanw/}.

\bibitem[{\citenamefont{Boyom}(2002)}]{Boyom}
\bibinfo{author}{\bibfnamefont{M.~N.} \bibnamefont{Boyom}}
  (\bibinfo{year}{2002}), \eprint{arXiv:math.DG/0202259}.

\bibitem[{\citenamefont{Bojowald et~al.}(2003)\citenamefont{Bojowald, Kotov,
  and Strobl}}]{BKS}
\bibinfo{author}{\bibfnamefont{M.}~\bibnamefont{Bojowald}},
  \bibinfo{author}{\bibfnamefont{A.}~\bibnamefont{Kotov}}, \bibnamefont{and}
  \bibinfo{author}{\bibfnamefont{T.}~\bibnamefont{Strobl}}
  (\bibinfo{year}{2003}), \bibinfo{note}{in preparation}.

\bibitem[{\citenamefont{Sullivan}(1977)}]{Sullivan}
\bibinfo{author}{\bibfnamefont{D.}~\bibnamefont{Sullivan}},
  \bibinfo{journal}{Inst. des Haut Etud. Sci. Pub. Math}
  \textbf{\bibinfo{volume}{47}}, \bibinfo{pages}{269} (\bibinfo{year}{1977}).

\bibitem[{\citenamefont{Izquierdo}(1999)}]{Izquierdo}
\bibinfo{author}{\bibfnamefont{J.~M.} \bibnamefont{Izquierdo}},
  \bibinfo{journal}{Phys. Rev.} \textbf{\bibinfo{volume}{D59}},
  \bibinfo{pages}{084017} (\bibinfo{year}{1999}), \eprint{hep-th/9807007}.

\bibitem[{\citenamefont{Schaller and Strobl}(1994{\natexlab{c}})}]{PLB}
\bibinfo{author}{\bibfnamefont{P.}~\bibnamefont{Schaller}} \bibnamefont{and}
  \bibinfo{author}{\bibfnamefont{T.}~\bibnamefont{Strobl}},
  \bibinfo{journal}{Phys. Lett.} \textbf{\bibinfo{volume}{B337}},
  \bibinfo{pages}{266} (\bibinfo{year}{1994}{\natexlab{c}}),
  \eprint[http://arXiv.org/abs]{hep-th/9401110}.

\bibitem[{\citenamefont{Teitelboim}(1983)}]{teitelboim:1983ux}
\bibinfo{author}{\bibfnamefont{C.}~\bibnamefont{Teitelboim}},
  \bibinfo{journal}{Phys. Lett.} \textbf{\bibinfo{volume}{B126}},
  \bibinfo{pages}{41} (\bibinfo{year}{1983}).

\bibitem[{\citenamefont{Jackiw}(1984)}]{jackiw84}
\bibinfo{author}{\bibfnamefont{R.}~\bibnamefont{Jackiw}}, in
  \emph{\bibinfo{booktitle}{Quantum Theory of Gravity. Essays in Honor of the
  60th Birthday of Bryce S. DeWitt}}, edited by
  \bibinfo{editor}{\bibfnamefont{S.}~\bibnamefont{Christensen}}
  (\bibinfo{publisher}{Hilger, Bristol}, \bibinfo{year}{1984}), pp.
  \bibinfo{pages}{403--420}.

\bibitem[{\citenamefont{Kl{\"o}sch and Strobl}(1998)}]{kinks}
\bibinfo{author}{\bibfnamefont{T.}~\bibnamefont{Kl{\"o}sch}} \bibnamefont{and}
  \bibinfo{author}{\bibfnamefont{T.}~\bibnamefont{Strobl}},
  \bibinfo{journal}{Phys. Rev.} \textbf{\bibinfo{volume}{D57}},
  \bibinfo{pages}{1034} (\bibinfo{year}{1998}), \eprint{arXiv:gr-qc/9707053}.

\bibitem[{\citenamefont{Achucarro and Townsend}(1986)}]{Achucarro}
\bibinfo{author}{\bibfnamefont{A.}~\bibnamefont{Achucarro}} \bibnamefont{and}
  \bibinfo{author}{\bibfnamefont{P.~K.} \bibnamefont{Townsend}},
  \bibinfo{journal}{Phys. Lett.} \textbf{\bibinfo{volume}{B180}},
  \bibinfo{pages}{89} (\bibinfo{year}{1986}).

\bibitem[{\citenamefont{Witten}(1988)}]{Witten2+1}
\bibinfo{author}{\bibfnamefont{E.}~\bibnamefont{Witten}},
  \bibinfo{journal}{Nucl. Phys.} \textbf{\bibinfo{volume}{B311}},
  \bibinfo{pages}{46} (\bibinfo{year}{1988}).

\bibitem[{\citenamefont{Grumiller
  et~al.}(2003{\natexlab{b}})\citenamefont{Grumiller, Kummer, Mignemi, Strobl,
  and Vassilevich}}]{5peopleinprep}
\bibinfo{author}{\bibfnamefont{D.}~\bibnamefont{Grumiller}},
  \bibinfo{author}{\bibfnamefont{W.}~\bibnamefont{Kummer}},
  \bibinfo{author}{\bibfnamefont{S.}~\bibnamefont{Mignemi}},
  \bibinfo{author}{\bibfnamefont{T.}~\bibnamefont{Strobl}}, \bibnamefont{and}
  \bibinfo{author}{\bibfnamefont{D.}~\bibnamefont{Vassilevich}}
  (\bibinfo{year}{2003}{\natexlab{b}}), \bibinfo{note}{in preparation}.

\bibitem[{\citenamefont{D{\"u}chting and Strobl}(1999)}]{Norbert}
\bibinfo{author}{\bibfnamefont{N.}~\bibnamefont{D{\"u}chting}}
  \bibnamefont{and} \bibinfo{author}{\bibfnamefont{T.}~\bibnamefont{Strobl}}
  (\bibinfo{year}{1999}), \bibinfo{note}{unpublished notes}.

\bibitem[{\citenamefont{Roytenberg}(1999)}]{Roytenbergdiss}
\bibinfo{author}{\bibfnamefont{D.}~\bibnamefont{Roytenberg}}, Ph.D. thesis,
  \bibinfo{school}{UC Berkeley} (\bibinfo{year}{1999}),
  \eprint[http://arXiv.org/abs]{math.DG/9910078}.

\bibitem[{\citenamefont{Alekseev and Strobl}(2003)}]{AS}
\bibinfo{author}{\bibfnamefont{A.}~\bibnamefont{Alekseev}} \bibnamefont{and}
  \bibinfo{author}{\bibfnamefont{T.}~\bibnamefont{Strobl}}
  (\bibinfo{year}{2003}), \bibinfo{note}{in preparation}.

\end{thebibliography}

\end{document}